\documentstyle[12pt,epsfig]{article}
\def\theequation{\arabic{section}.\arabic{equation}}
\newcommand{\alt}{\,\rlap{\lower 3.5 pt \hbox{$\mathchar \sim$}} \raise 1pt
 \hbox {$<$}\,}

\catcode`@=11
\newcount\@tempcntc
\def\@citex[#1]#2{\if@filesw\immediate\write\@auxout{\string\citation{#2}}\fi
  \@tempcnta\z@\@tempcntb\m@ne\def\@citea{}\@cite{\@for\@citeb:=#2\do
    {\@ifundefined
       {b@\@citeb}{\@citeo\@tempcntb\m@ne\@citea\def\@citea{,}{\bf ?}\@warning
       {Citation `\@citeb' on page \thepage \space undefined}}%
    {\setbox\z@\hbox{\global\@tempcntc0\csname b@\@citeb\endcsname\relax}%
     \ifnum\@tempcntc=\z@ \@citeo\@tempcntb\m@ne
       \@citea\def\@citea{,}\hbox{\csname b@\@citeb\endcsname}%
     \else
      \advance\@tempcntb\@ne
      \ifnum\@tempcntb=\@tempcntc
      \else\advance\@tempcntb\m@ne\@citeo
      \@tempcnta\@tempcntc\@tempcntb\@tempcntc\fi\fi}}\@citeo}{#1}}
\def\@citeo{\ifnum\@tempcnta>\@tempcntb\else\@citea\def\@citea{,}%
  \ifnum\@tempcnta=\@tempcntb\the\@tempcnta\else
   {\advance\@tempcnta\@ne\ifnum\@tempcnta=\@tempcntb \else \def\@citea{--}\fi
    \advance\@tempcnta\m@ne\the\@tempcnta\@citea\the\@tempcntb}\fi\fi}
\catcode`@=12

\begin{document}

\title{\vskip-3cm{\baselineskip14pt
\centerline{\normalsize\hfill MPI/PhT/97--050}
\centerline{\normalsize\hfill hep--ph/9709361}
\centerline{\normalsize\hfill September 1997}
}
\vskip1.5cm
Loop Effects of Exotic Leptons on Vector-Boson Pair Production at $e^+e^-$
Colliders
}
\author{{\sc K.-P.O. Diener, B.A. Kniehl, and A. Pilaftsis}\\
Max-Planck-Institut f\"ur Physik (Werner-Heisenberg-Institut),\\
F\"ohringer Ring 6, 80805 Munich, Germany}
\date{}
\maketitle

\thispagestyle{empty}

\begin{abstract}

We study quantum effects of exotic heavy leptons on the production of two
vector bosons in $e^+e^-$ annihilation.
We present closed analytic expressions for vector-boson self-couplings and
differential cross sections, within the framework of two favourable
new-physics scenarios: (i) the fourth-generation Majorana-neutrino model and
(ii) an  E$_6$-inspired model with sequential mirror isodoublets.
We constrain these models by requiring that their contributions to the
oblique electroweak parameters be compatible with a recent global fit to
high-precision data.
We then systematically analyze the loop-induced deviations in the LEP2 cross
section of $e^+e^-\to W^+W^-$ predicted by the models thus confined.
\medskip

\end{abstract}

\newpage

\section{Introduction}

Neutrinos are the most weakly coupled  particles observed in nature, a
fact  that makes their   detailed experimental study rather difficult.
Understanding the  underlying properties of  these massless  or almost
massless  neutral fermions will  definitely shed light  on a number of
fundamental questions in particle  and astro-particle physics.  In the
minimal  Standard Model (SM), the  neutrinos  are strictly massless by
construction, due to the    absence of right-handed    neutrino states.
However, all other fermionic  matter, {\it i.e.}, the charged  leptons
and quarks, require the  existence of right-handed fields.  How should
one  understand theoretically this asymmetry  of the fermionic degrees
of freedom in  the SM?
Another  long-standing  puzzle is  related  to the energy
deficit  of the solar and  atmospheric neutrinos.  A possible solution
to all   these problems may  be  achieved   by means of   the  see-saw
mechanism \cite{YAN}, which   requires the existence of   right-handed
neutrino fields  in  addition to the   left-handed  ones of the minimal SM.
Then, very large Majorana masses can be present in the extended SM
Lagrangian,  and,   together  with  Dirac  terms  of order  of
charged-lepton or quark  masses, form the  entries of the see-saw mass
matrix.   Diagonalization of the  see-saw mass  matrix naturally gives
rise to  non-zero, but very   small neutrino masses,  in agreement  with
experiment, as  well as  to  ultra-heavy neutrinos not yet discovered.
Furthermore, a   possible explanation  of   the atmospheric  and solar
neutrino problems  may  be based  on the  Mikheyev-Smirnov-Wolfenstein
mechanism  \cite{MSW},  which also  requires that, contrary to the SM, the
known   neutrinos  be   massive.    Finally,   reconciliation  of  the
atmospheric, solar, and Large Scale Neutrino Detector (LSND) anomalies
is only possible if four low-mass neutrinos exist \cite{fournus}.
The latter  may be regarded as a
strong indication in favour of fourth-generation extensions of the SM.

If all neutrino problems can indeed be resolved by the presence of new
heavy  neutral   leptons,  one   has then  to    investigate  possible
consequences emanating from these exotic particles.
There are  strict experimental bounds from direct searches at LEP1 and, to 
a lesser extend, at LEP2 for new heavy fermions  with appreciable couplings to
the $Z$ boson. However, even if  heavy particles are not directly accessible at
present  energies, they  can  still leave  their imprints by  inducing
noticeable   deviations at the  quantum  level through the oblique electroweak
parameters, which have now  been tightly constrained by
LEP1 measurements. Such heavy particles may also have an observable impact on
the triple-gauge-boson couplings, which are directly accessible at LEP2.

In this paper, we study loop effects of exotic heavy leptons on the cross 
sections of the following four processes:
(i) $e^+e^-\to\gamma\gamma$; (ii) $e^+e^-\to\gamma Z$;
(iii) $e^+e^-\to ZZ$; and (iv) $e^+e^-\to W^+W^-$.
To that end, we consider two favourable models, which are renormalizable.
In the first model, which was
first discussed by Hill and Paschos (HP) \cite{HP}, the fermionic sector of
the  SM is  extended by   adding one  sequential  weak isodoublet, one
right-handed  neutrino, and  one  right-handed  charged lepton.   After
spontaneous symmetry breaking, the left-handed neutrinos couple to the
right-handed neutrino with  Dirac masses  proportional to the  charged
lepton masses,  whereas the right-handed  neutrino develops a Majorana
mass at the electroweak scale.  The HP model can naturally predict a mass
hierarchy between the  different neutrino species.   Since the charged
lepton  of  the fourth  generation  must be heavy for phenomenological
reasons, the fourth doublet neutrino  then turns out   to be heavy  as
well.  Furthermore, inter-family mixing  between the three generations
and the fourth  one is suppressed  by the usual see-saw relations and
can thus be  safely neglected. To   render the HP model anomaly-free,
however, one  should include one additional quark isodoublet and one
pair of up-type and down-type right-handed quarks. 

The second   favourable model that  could potentially  allow for large
loop effects of exotic leptons is due to Ma  and Roy (MR) \cite{MR}. As was
argued in Ref.~\cite{MR},  the MR model may be a  viable low-energy limit  of
E$_6$ unified  theories.  The MR model is slightly more economical than the
HP model,  since  it only contains  two  colourless doublets  with
opposite  weak isospins and one isosinglet   neutrino in addition to
the SM  field content.  Because of this  mirror assignment  of the two
weak  isodoublets, the MR model  is anomaly free  by  itself; the quark
sector does not need to be extended.  Apart from Dirac masses, the isosinglet
neutrino can   introduce a lepton-number-violating  Majorana mass into
the Lagrangian.  As we shall see, the  two mirror isodoublets can form
a singlet mass term for  the heavy charged lepton.  Among other things,
the MR  model leads to both left-  and right-handed couplings  of the $W$
boson to the heavy charged lepton and the heavy neutrinos.

Recently, an analysis of heavy-lepton effects on the reaction
$e^+e^-\to  W^+W^-$ in  the  context of the HP model has been reported
\cite{Japan}.
We have checked that our results agree with those of Ref.~\cite{Japan}.
The salient difference between our analysis and that of Ref.~\cite{Japan}
resides in the  fact  that our numerical predictions  are obtained after
imposing the constraints on the parameters of the HP model that come from
experimental bounds on the oblique electroweak parameters and from direct
searches for new heavy leptons.
To our knowledge, similar studies have not yet been performed for the
production of neutral vector bosons or for the MR model.

This paper is organized as follows. In Section~\ref{sec:models},
we describe  the basic low-energy structure  of the  HP
and MR  scenarios. Section~\ref{sec:vvv} contains,
in analytic form, the resulting one-loop corrections to the cross sections of
processes~(i)--(iv).
Technical details of the
calculation  are relegated to the  Appendices.  In Section~\ref{sec:discus},
we present numerical predictions, relevant for LEP2, for the loop effects due
to exotic leptons within the context of  the   HP and   MR  models, after the
parameter  spaces of these  models  have  been constrained by imposing limits
derived from oblique electroweak parameters and direct searches.  
Our conclusions are summarized in Section~\ref{sec:concl}.

\setcounter{equation}{0}
\section{Outline of the models\label{sec:models}}

In the following, we present a brief outline of the HP and MR models.
Both models
are based on the SM  gauge group SU(2)$\times$U(1), but contain
new isodoublet and isosinglet fields.   The resulting gauge interactions
are different for the two models, which in  turn may lead to different
phenomenological consequences.

\subsection{Hill-Paschos model}

In the HP model \cite{HP}, the lepton and quark isodoublets $(\nu',E)_L$ and
$(T,B)_L$ as well as the four right-handed fields $E_R$, $N'_R$, $T_R$, and
$B_R$ are added to the SM.
To avoid tight limits coming from the observed sector, we assume the absence
of any mixing between the ordinary matter and the new fields introduced in
the model.
The heavy quarks $T$ and $B$ are needed to cancel the triangle anomalies.
They are usually considered to be mass-degenerate, {\it i.e.},
$m_T = m_B$, in order to avoid large contributions to the electroweak
$\rho$ parameter.

In the lepton  sector, the HP  model  predicts a heavy  charged lepton
with  mass $m_E$, which is to  be constrained by electroweak radiative
corrections.  Moreover, the  isosinglet neutrino  $N'_R$ admits the
presence of the  gauge-invariant Majorana mass term $m_M \bar{N}'^C_RN'_R$,
so that the mass Lagrangian of the neutrinos has the non-standard  form
\begin{equation}
\label{LmassHP}
{\cal L}_M^\nu = - \frac{1}{2} 
\left(\bar{N}'_L,  \bar{N}'^C_R\right)
  \left(
    \begin{array}{cc} 0 & m_D \\
      m_D & m_M 
    \end{array}
  \right)
\left( \begin{array}{c} N'^C_L\\ N'_R \end{array} \right) + {\rm H.c.}
\end{equation}
The $2\times2$ neutrino mass  matrix in Eq.~(\ref{LmassHP}), which we call
$M^\nu$, can always  be
diagonalized  through   a  unitary  transformation as $U^T   M^\nu  U =
\widehat{M}^\nu$. After the  diagonalization of  ${\cal L}_M^\nu$,  we
obtain two heavy  Majorana  neutrinos, $\nu$  and  $N$.  In  the limit
$m_D\ll  m_M$,  $\nu$ is   predominantly  isodoublet, whereas  $N$  is
isosinglet.  In the   HP model,  $m_D\sim    m_E$ and  $m_M$ is of the order
of the electroweak scale $G_F^{1/2}$, {\it i.e.}, $m_M\alt1$~TeV. 

The interactions of the two Majorana neutrinos $n_1\equiv\nu$ and
$n_2\equiv N$ with the $W$ and $Z$ bosons and the charged lepton $E$ is
described by the Lagrangian
\begin{eqnarray}
\label{LintHP}
{\cal L}_{\rm int}&=& 
\frac{e}{\sqrt{2}s_w} W^-_{\mu} \sum_{i=1,2} B_{Ei}  
\bar{E}\gamma^{\mu} {\rm P}_- n_i  + {\rm H.c.}
\nonumber\\
 & &+  \frac{e}{4c_w s_w } Z_{\mu} \sum_{i,j=1,2}  
\bar{n}_i \gamma^{\mu}
\left( i \Im m C_{ij} - \gamma_5 {\rm Re} C_{ij}\right) n_j ,
\end{eqnarray}
where ${\rm P}_\pm=(1\pm\gamma_5)/2$ are the right/left-handed helicity 
projectors, $e$ is the electron charge magnitude, $c_w^2=1-s_w^2=M_W^2/M_Z^2$,
and, according to the conventions of Ref.~\cite{ZPC},
$B_{Ei} =  U_{1i}^*$ and $C_{ij}  = U_{1i} U^*_{1j}$.
The mixing $B_{Ei}$ and $C_{ij}$ may be expressed in terms
of  the physical  Majorana    masses $m_\nu$ and  $m_N$  as \cite{KP}
\begin{eqnarray}
\label{MixHP}
B_{E\nu} &=& \sqrt{\frac{m_N}{m_\nu + m_N}},\qquad
B_{EN}\ = i \sqrt{\frac{m_\nu}{m_\nu + m_N}},
\nonumber\\
C_{\nu\nu} &=& \frac{m_N}{m_\nu}C_{NN} = \frac{m_N}{m_\nu + m_N},\qquad
C_{\nu N} = -C_{N\nu} = i \frac{\sqrt{m_\nu m_N}}{m_\nu + m_N} .
\end{eqnarray}
For later convenience, we rewrite the interaction Lagrangian (\ref{LintHP}) as
\begin{eqnarray}
\label{LGintHP}
{\cal L}_{\rm int}&=& e W^-_{\mu} \sum_{i=1,2} 
\bar{E}\gamma^{\mu} 
\left(g^+_{En_iW^-} {\rm P}_+ + g^-_{En_iW^-} {\rm P}_-\right) n_i + {\rm H.c.}
\nonumber\\
& & + e  Z_{\mu} \sum_{i,j=1,2}
\bar{n}_i \gamma^{\mu}\left( g^+_{n_in_jZ} {\rm P}_+ +  
g^-_{n_in_jZ} {\rm P}_- \right)n_j ,
\end{eqnarray}
where
\begin{eqnarray}
   \label{gparHP}
  g^+_{En_iW^-} &=& 0, \qquad  g^-_{En_iW^-} =  
                                    \frac{B_{Ei}}{\sqrt{2}s_w } ,
\nonumber\\ 
g^+_{n_in_jZ} &=&- \frac{C_{ij}^*}{4c_w  s_w},\qquad
  g^-_{n_in_jZ} = \frac{C_{ij}}{4c_w  s_w}  .
\end{eqnarray}
The remaining interactions of the gauge bosons with the quarks and
charged leptons are of the SM type, with couplings
\begin{eqnarray}
\label{gparSM}
g_{BTW^-}^+&=&0,\qquad
g_{BTW^-}^-=\frac{1}{\sqrt2s_w},\nonumber\\
g_{FFZ}^+&=&-\frac{s_wQ_F}{c_w},\qquad
g_{FFZ}^-=\frac{T_F}{c_ws_w}-\frac{s_wQ_F}{c_w},\nonumber\\
g_{FF\gamma}^\pm&=&-Q_F,
\end{eqnarray}
where $Q_F$ and $T_F$ are the electric charge and the third component of the
weak isospin of $F=E,T,B$, respectively.

\subsection{Ma-Roy model}

In the MR model \cite{MR},  only the  lepton sector  of the SM is extended.
Specifically, two colorless  doublets $(N_1,E)_L$ and $(E^C,N_2)_L$ and a
colorless isosinglet ${N_3}_L$   are added.  Similarly to the HP
model, we  assume that the  new fields  do not  mix  with the observed
leptons. Since the two new isodoublets have opposite hypercharges, the
triangle anomalies cancel, so that there is
no need to extend the quark sector as well. As is argued in Ref.~\cite{MR},
such an   anomaly-free representation may be   the low-energy limit of
unified E$_6$ models.

In addition to the Majorana mass term $m_M\bar{N}^C_{3L}N_{3L}$, the
two lepton isodoublets can form the gauge-invariant isosinglet mass term of 
the type
\begin{displaymath}
m_E\left[\left(E^C_L\right)^T C^{-1} E_L - N^T_{2L} C^{-1} N_{1L}\right] +
m_E\left[E_L^T C^{-1} E^C_L - N^T_{1L} C^{-1} N_{2L}\right] .
\end{displaymath}
In this  way, the charged lepton $E$ of the MR model acquires a
SU(2)$\times$U(1)-invariant mass term.
Furthermore, after spontaneous symmetry breaking, the neutrino mass Lagrangian
takes the form
\begin{equation}
   \label{LmassMR}
{\cal L}_M^\nu = - \frac{1}{2} 
\left(\bar{N}^C_{1L}, \bar{N}^C_{2L}, \bar{N}^C_{3L} \right)
  \left(
  \begin{array}{ccc}
    0 & -m_E & m_1 \\  -m_E & 0  & m_2 \\ m_1 &  m_2 & m_M \end{array}
  \right) \left( \begin{array}{c} N_{1L}\\ N_{2L}\\ N_{3L} \end{array}
\right) + {\rm H.c.}
\end{equation}
For simplicity,  we assume that  the $3\times  3$ neutrino mass matrix,
$M^\nu$, is  real.   Obviously, the   MR  model predicts  three  heavy
Majorana neutrinos, which are denoted by $n_1$, $n_2$, and $n_3$.  The
diagonalization of  $M^\nu$ as well as  the explicit analytic  form of
the  three mass eigenvalues  and  eigenvectors are  given in
Appendix~\ref{eigenvals}.

The interactions of  the  $Z$  and $W$  bosons with the heavy   Majorana
neutrinos $n_i$ ($i=1,2,3$)  and the charged  lepton $E$ are determined
by the Lagrangian
\begin{eqnarray}
 \label{LintMR}
{\cal L}_{\rm int} &=& \frac{e}{2\sqrt{2}s_w } 
                     W_\mu^- \sum_{i=1}^3\bar E\gamma^\mu 
          \left[\left(U_{1i} - U_{2i}^*\right) - \gamma_5
 \left(U_{1i} +U_{2i}^*\right)            \right] n_i  +  {\rm H.c.}
\nonumber\\
         & & + \frac{e}{4c_w s_w } Z_\mu \left[ \sum_{i,j=1}^3
 \bar{n}_i \gamma^\mu\left(i\Im m C_{ij} - \gamma_5 {\rm Re} C_{ij}\right) n_j 
              + 2(2s^2_w -1)\bar{E} \gamma^\mu E \right] ,
\end{eqnarray}
where $C_{ij}  = U^*_{1i}U_{1j}  -  U^*_{2i}U_{2j}$.  Notice that  the
coupling of the charged lepton $E$  to the $Z$ boson is purely
vectorial, whereas the $W$ boson couples  with both chiralities to the $E$
and $n_i$ fields. This makes the MR scenario very distinctive from the
HP model.  Using a parameterization analogous to Eq.~(\ref{LGintHP}),
complemented by the $Z\bar{E}E$ interaction Lagrangian
\begin{equation}
   \label{ZEE}
{\cal L}_{int}^Z = e  Z_\mu  \bar{E} \gamma^\mu 
(g^+_{EEZ} {\rm P}_+ + g^-_{EEZ}{\rm P}_-) E,
\end{equation}
the couplings of the model read
\begin{eqnarray}
\label{gparMR}
g_{En_iW^-}^+&=&-\frac{U_{2i}^*}{\sqrt2s_w},\qquad 
g_{En_iW^-}^-=\frac{U_{1i}}{\sqrt2s_w},\nonumber\\
g_{n_in_jZ}^+&=&-\frac{C_{ij}^*}{4c_ws_w},\qquad
g_{n_in_jZ}^-=\frac{C_{ij}}{4c_ws_w},\qquad
g_{EEZ}^\pm=\frac{2s^2_w-1}{2c_ws_w}.
\end{eqnarray}

\setcounter{equation}{0}
\section{Analytic results\label{sec:vvv}}

In this section, we calculate the quantum corrections to the cross sections of 
processes~(i)--(iv) specified in the Introduction which are induced by the
exotic heavy fermions of the HP and MR models.
As a reference, we also list the tree-level results.
For the reader's convenience, technical details such as the definitions of the
relevant matrix elements, one-loop functions, and renormalization constants
are relegated to Appendices~\ref{stdmatrix}, \ref{oneloop}, and \ref{renorm},
respectively.

We denote the four-momenta of $e^+$, $e^-$, and the two produced vector bosons,
$V_1$ and $V_2$, by $p_+$, $p_-$, $k_1$, and $k_2$, and define the Mandelstam 
variables as $s=(p_++p_-)^2$, $t=(p_+-k_1)^2$, and $u=(p_+-k_2)^2$.
Neglecting the electron mass, we have $s+t+u=M_1^2+M_2^2$, where $M_1$ and 
$M_2$ are the masses of $V_1$ and $V_2$, respectively.
In this limit, also the $s$-channel contributions due to Higgs-boson exchanges
vanish.
Because each of the processes~(i)--(iv) has more than one tree-level diagram,
it proves convenient to present the analytic results in terms of helicity
amplitudes.
In the centre-of-mass (c.m.) system, the electron and positron have opposite
helicities, so that one helicity label $\kappa=\pm$ for left/right-handed
electron helicity suffices.
Calling the c.m.\ helicities of $V_1$ and $V_2$, $\lambda_1$ and $\lambda_2$,
the differential cross section may be cast into the generic form
\begin{equation}
\label{dsigma}
\frac{{\rm d}\sigma}{{\rm d}\Omega}=
\frac{\lambda^{1/2}(s,M^2_1,M^2_2)}{64\pi^2s^2}\,\frac{1}{4}
\sum_{\kappa,\lambda_1,\lambda_2} 
\left|{\cal M}^\kappa(\lambda_1,\lambda_2,s,t)\right|^2,
\end{equation}
where $\lambda(x,y,z)=x^2+y^2+z^2-2(xy+yz+zx)$ is the K\"all\'en function
and the factor $1/4$ stems from the average over the initial spins.
In the following, we suppress the arguments of the helicity amplitudes
${\cal M}^\kappa$ if confusion is impossible.
It is convenient to express the amplitudes ${\cal M}^\kappa$ as linear
combinations of the standard matrix elements ${\cal M}_i^\kappa$
($i=0,\ldots,9$) listed in Eq.~(\ref{StandardME}), which have well-known
helicity representations \cite{Sack}.
In addition, it is useful to introduce the following combinations pertinent
to $s$-, $t$-, and $u$-channel exchange: 
\begin{eqnarray}
\label{Mstu}
{\cal M}^\kappa_s&=&2\frac{e^2}{s} 
\left({\cal M}^\kappa_1-{\cal M}^\kappa_2-{\cal M}^\kappa_3\right),
\nonumber\\ 
{\cal M}^\kappa_t&=&-\frac{e^2}{t}{\cal M}^\kappa_0,
\nonumber\\
{\cal M}^\kappa_u&=&-\frac{e^2}{u}\left({\cal M}^\kappa_0+
2{\cal M}^\kappa_1-2{\cal M}^\kappa_2-2{\cal M}^\kappa_3\right).
\end{eqnarray}

In the one-loop approximation, each helicity amplitude ${\cal M}^\kappa$ is
expanded in the fine-structure constant $\alpha=e^2/(4\pi)$ as
${\cal M}^\kappa={\cal M}_{\rm Born}^\kappa+\delta{\cal M}^\kappa$,
where ${\cal M}_{\rm Born}^\kappa$ and $\delta{\cal M}^\kappa$ are the
tree-level and one-loop contributions, respectively.
In turn, $\delta {\cal M}^\kappa$ receives contributions from diagrams
containing self-energy corrections, vertex corrections, and counterterm
insertions, {\it i.e.},
$\delta{\cal M}^\kappa=\delta{\cal M}_S^\kappa
+\delta{\cal M}_V^\kappa+\delta{\cal M}_C^\kappa$.
Consequently, through order $\alpha^3$, the differential cross
section~(\ref{dsigma}) may be written as
\begin{equation}
\label{dsigma1}
\frac{{\rm d}\sigma}{{\rm d}\Omega}= 
   \left( \frac{{\rm d}\sigma}{{\rm d}\Omega}\right)_{\rm Born} + 
    \frac{\lambda^{1/2}}{128\pi^2s^2} \sum_{\kappa,\lambda_1,\lambda_2}  
     {\rm Re}\left[\left({\cal M}_{\rm Born}^\kappa\right)^*
\delta{\cal M}^\kappa\right].
\end{equation}

First, we present ${\cal M}_{\rm Born}^\kappa$ for processes~(i)--(iv).
As is well known \cite{Sack}, we have
\begin{eqnarray}
\label{Born}
{\cal M}_{\rm Born}^\kappa(\gamma\gamma)&=&
{\cal M}_t^\kappa+{\cal M}_u^\kappa,
\nonumber\\
{\cal M}_{\rm Born}^\kappa(\gamma Z)&=& 
g_{eeZ}^\kappa\left({\cal M}_t^\kappa+{\cal M}_u^\kappa\right),
\nonumber\\
{\cal M}_{\rm Born}^\kappa(ZZ)&=&
\left(g_{eeZ}^\kappa\right)^2\left({\cal M}_t^\kappa+{\cal M}_u^\kappa\right),
\nonumber\\
{\cal M}_{\rm Born}^+(WW)&=&
{\cal M}_s^+\left(1-g_{eeZ}^+\frac{c_w}{s_w}\,\frac{s}{s-M^2_Z}\right),
\nonumber\\ 
{\cal M}_{\rm Born}^-(WW)&=&
{\cal M}_s^-\left(1-g_{eeZ}^-\frac{c_w}{s_w}\,\frac{s}{s-M^2_Z}\right)
+\frac{{\cal M}_t^-}{2s_w^2},
\end{eqnarray}
where $g_{eeZ}^+=s_w/c_w$ and $g_{eeZ}^-=(2s_w^2-1)/(2s_wc_w)$ are the SM
couplings of the electron to the $Z$ boson in the notation of
Eq.~(\ref{gparSM}).

Next, we present the contributions to the $\gamma\gamma$, $ZZ$, and $WW$
self-energies and to the $\gamma Z$ mixing amplitude induced by the exotic
fermions of the HP and MR models.
These will enter the expressions for $\delta{\cal M}_S^\kappa$ and
$\delta{\cal M}_C^\kappa$.
Instead of listing all individual contributions separately, we present a
generic expression, which depends on the coupling parameters defined in
Eqs.~(\ref{gparHP}), (\ref{gparSM}), and (\ref{gparMR}).
Since all the gauge bosons involved in processes~(i)--(iv) couple to conserved
currents, only the transverse parts of the above-named vacuum polarizations,
$\Pi_T^{V_1V_2}$ with $V_1V_2=\gamma\gamma,\gamma Z,ZZ,WW$, need be
calculated.
In dimensional regularization with $D$ space-time dimensions, we have
\begin{eqnarray}
\label{PiT}
\Pi^{V_1V_2}_T(p^2)&=&\frac{1}{D-1}
\left(g^{\mu\nu}-\frac{p^\mu p^\nu}{p^2}\right)\sum_{(f_i,f_j)}
\frac{\alpha}{4\pi}N_c^fN_S^f\frac{(2\pi\mu)^{4-D}}{i\pi^2}\int{\rm d}^Dq\,
{\rm Tr}\left[\gamma_\nu\left(g_{ijV_2}^+{\rm P}_+
\right.\vphantom{\frac{1}{{\not\! q}+{\not\! p}-m_j+i\varepsilon}}\right.
\nonumber\\
&&\left.\left.+g_{ijV_2}^-{\rm P}_-\right)
\frac{1}{{\not\! q}+{\not\! p}-m_j+i\varepsilon}\gamma_\mu
\left(g_{jiV_1}^+{\rm P}_++g_{jiV_1}^-{\rm P}_-\right)
\frac{1}{{\not\! q}-m_i+i\varepsilon}\right]\nonumber\\
&=&\sum_{(f_i,f_j)}\pi\alpha N_c^fN_S^f\left\{
\left(g_{jiV_1}^+g_{ijV_2}^++g_{jiV_1}^-g_{ijV_2}^-\right)
\left[\Pi_V(p^2,m_i,m_j)+\Pi_V(p^2,m_i,-m_j)\right]\right.
\nonumber\\
&&\left.+
\left(g_{jiV_1}^+g_{ijV_2}^-+g_{jiV_1}^-g_{ijV_2}^+\right)
\left[\Pi_V(p^2,m_i,m_j)-\Pi_V(p^2,m_i,-m_j)\right]\right\},
\end{eqnarray}
where $p$ is the external four-momentum,
the sum runs over all possible pairings $(f_i,f_j)$ of the exotic fermions,
$N_c^f=1$ (3) for leptons (quarks),
we have introduced the short-hand notations $m_i=m_{f_i}$ and 
$g_{ijV}^\pm=g_{f_if_jV}^\pm$,
and the $\Pi_V$ function is defined in Eq.~(\ref{eqpiv}).
The combinatorial factor $N_S^f$ accounts for the Majorana properties of the
heavy neutrinos; it takes the value $N_S^f=2$ if $f_i$ and $f_j$ are both
Majorana neutrinos, while $N_S^f=1$ otherwise.
The unphysical 't~Hooft mass scale, $\mu$, is introduced to keep the coupling
constants dimensionless; it cancels along with the ultraviolet singularities
upon renormalization.
Equation~(\ref{PiT}) generalizes the results for the HP model found in 
Refs.~\cite{KP,Hansi}.

With the help of Eq.~(\ref{PiT}), we may present $\delta{\cal M}_S^\kappa$ for
processes~(i)--(iv) as
\begin{eqnarray}
\label{dMS}
\delta{\cal M}_S^\kappa(\gamma\gamma)&=&
\delta{\cal M}_S^\kappa(\gamma Z)=\delta{\cal M}_S^\kappa(ZZ)=0,
\nonumber\\
\delta{\cal M}_S^\kappa(WW)&=&{\cal M}_s^\kappa\left[
-\frac{\Pi_T^{\gamma\gamma}(s)}{s}
+\left(\frac{c_w}{s_w}-g_{eeZ}^\kappa\right)
\frac{\Pi_T^{\gamma Z}(s)}{s-M_Z^2}
+g_{eeZ}^\kappa\frac{c_w}{s_w}\,\frac{s\Pi_T^{ZZ}(s)}{\left(s-M_Z^2\right)^2}
\right].
\end{eqnarray}
The first line of Eq.~(\ref{dMS}) reflects the fact that 
processes~(i)--(iii) do not involve virtual vector bosons at tree level.

The mass and coupling counterterms as well as the wave-function
renormalization constants that are relevant for our analysis can all be 
expressed in terms of the $\Pi_T^{V_1V_2}$ functions of Eq.~(\ref{PiT}) and
their derivatives with respect to $p^2$.
The corresponding relations are summarized for the on-shell renormalization
scheme in Eq.~(\ref{eqos}).
The counterterm amplitudes $\delta{\cal M}_S^\kappa$ for the 
processes~(i)--(iv) emerge from the respective Born amplitudes
$\delta{\cal M}_{\rm Born}^\kappa$ of Eq.~(\ref{Born}) by scaling the
couplings and masses and by including the appropriate wave-function 
renormalizations.
The appropriate relations between the bare and renormalized parameters and
fields are collected in Eq.~(\ref{eqct}).
In this way, we obtain
\begin{eqnarray}
\label{CT}
\delta{\cal M}_C^\kappa(\gamma\gamma)&=&
{\cal M}_{\rm Born}^\kappa(\gamma\gamma)
\left(2\frac{\delta e}{e}
+\delta Z_{\gamma\gamma}
+g_{eeZ}^\kappa\delta Z_{Z\gamma}\right),
\nonumber\\
\delta{\cal M}_C^\kappa(\gamma Z)&=&  
{\cal M}_{\rm Born}^\kappa(\gamma Z)
\left(2\frac{\delta e}{e}
+\frac{\delta g_{eeZ}^\kappa}{g_{eeZ}^\kappa}
+\frac{\delta Z_{\gamma\gamma}}{2}
+\frac{g_{eeZ}^\kappa}{2}\delta Z_{Z\gamma}
+\frac{\delta Z_{\gamma Z}}{2g_{eeZ}^\kappa}
+\frac{\delta Z_{ZZ}}{2}\right),
\nonumber\\ 
\delta{\cal M}_C^\kappa(ZZ)&=&
{\cal M}_{\rm Born}^\kappa(ZZ)
\left(2\frac{\delta e}{e}
+2\frac{\delta g_{eeZ}^\kappa}{g_{eeZ}^\kappa}
+\frac{\delta Z_{\gamma Z}}{g_{eeZ}^\kappa}
+\delta Z_{ZZ}\right),
\nonumber\\
\delta{\cal M}_C^+(WW)&=&
{\cal M}_{\rm Born}^+(WW)
\left(2\frac{\delta e}{e}
+\delta Z_W
+\frac{\delta M_Z^2}{M_Z^2}\,\frac{s}{s-M_Z^2}\right),
\nonumber\\
\delta{\cal M}_C^-(WW)&=&
\frac{{\cal M}_s^-}{s-M_Z^2}
\left[\left(2\frac{\delta e}{e}+\delta Z_W\right)
\left(\frac{s}{2s_w^2}-M_Z^2\right)
+\left(\frac{1}{2s_w^2}-1\right)\delta M_Z^2\frac{s}{s-M_Z^2}
\right.\nonumber\\
&&\left.
-\frac{\delta s_w}{s_w^3}s\right]
+\frac{{\cal M}_t^-}{2s_w^2}
\left(2\frac{\delta e}{e}+\delta Z_W-2\frac{\delta s_w}{s_w}\right).
\end{eqnarray}
Notice that, since processes~(i)--(iii) do not involve virtual gauge bosons at
the tree level, their counterterm amplitudes only receive contributions from
coupling and wave-function renormalization.

Finally, we present the helicity amplitudes $\delta{\cal M}_V^\kappa$ 
containing the fermionic corrections to the triple-gauge-boson vertices in the
generic form:
\begin{eqnarray}
\label{3Vs}
\delta{\cal M}_V^\kappa
&=&\sum_{\kappa_1,\kappa_2,\kappa_3=\pm}
\sum_{B=\gamma,Z}\sum_{(f_i,f_j,f_k)}
\alpha^2N_c^fN_V^f\frac{g_{eeB}^\kappa}{s-M_B^2}
\left(g_{jkB}^{\kappa_3}g_{kiV_2}^{\kappa_2}g_{ijV_1}^{\kappa_1}
-g_{kjB}^{-\kappa_3}g_{ikV_2}^{-\kappa_2}g_{jiV_1}^{-\kappa_1}\right)
\nonumber\\
&&\times{\cal V}_{\kappa_1\kappa_2\kappa_3}^\kappa(\lambda_1,\lambda_2,s,t),
\end{eqnarray}
where the inner sum runs over all possible triplets $(f_i,f_j,f_k)$ of exotic 
loop fermions and
\begin{eqnarray}
\label{Vkappa}
{\cal V}_{\kappa_1\kappa_2\kappa_3}^\kappa&=&
\bar v(p_+)\gamma^\mu{\rm P}_\kappa u(p_-)
\varepsilon_1^\nu(k_1,\lambda_1)\varepsilon_2^\rho(k_2,\lambda_2)\\ 
&&\times\frac{(2\pi\mu)^{4-D}}{i\pi^2}\int{\rm d}^Dq\,{\rm Tr}\left(
\gamma_\mu{\rm P}_{\kappa_3}\frac{1}{\not\!q-\not\!k_2-m_k}
\gamma_\rho{\rm P}_{\kappa_2}\frac{1}{\not\!q-m_i}
\gamma_\nu{\rm P}_{\kappa_1}\frac{1}{\not\!q+\not\!k_1-m_j}\right).
\nonumber
\end{eqnarray}
The combinatorial factor $N_V^f$ accounts for the Majorana properties of the
heavy neutrinos; it takes the values $N_V^f=2^n$, where $n$ is the number of
$n_in_jZ$ couplings in the respective term of Eq.~(\ref{3Vs}).
The values of $g_{f_if_jV}^\kappa$ for the exotic leptons of the HP and MR
models are listed in Eqs.~(\ref{gparHP}) and (\ref{gparMR}), respectively;
the remaining SM-type couplings are given in Eq.~(\ref{gparSM}).
Analytic expressions for ${\cal V}_{\kappa_1\kappa_2\kappa_3}^\kappa$ in terms
of standard one-loop tensor integrals may be found in Eq.~(\ref{eqv}).
The first (second) term contained within the parentheses of Eq.~(\ref{3Vs})
stems from the direct (crossed) triangle diagram.
For $V_1V_2=W^+W^-$, the direct (crossed) triangle corresponds to the case 
when $f_i$ is down-type (up-type) and $f_j$ and $f_k$ are up-type (down-type).
If $V_1V_2=\gamma\gamma,\gamma Z,ZZ$, then, in the SM and the HP and MR 
models, only those parts of ${\cal V}_{\kappa_1\kappa_2\kappa_3}^\kappa$ in
Eq.~(\ref{eqv}) which carry an extra factor of $\kappa$ contribute to
Eq.~(\ref{3Vs}).
The SM version of Eq.~(\ref{3Vs}) agrees with Eq.~(5.23) of Ref.~\cite{Sack}
if we multiply $C_2^2$ in that equation by $k_2^2/k_1^2$.

It is interesting to note that, in contrast to the HP model, the MR model can
admit CP violation, if the mass parameters $m_1$, $m_2$, and $m_M$ are all
complex with different phases.
Here, we focus our attention on CP-conserving contributions to the
triple-gauge-boson couplings.
For a detailed discussion of CP-violating form-factors in Majorana-neutrino
models, we refer to Ref.~\cite{Cliff}.

\section{Numerical results and discussion\label{sec:discus}}

We are now in a position to explore the phenomenological implications of our 
results.
We focus our attention on LEP2 with c.m.\ energy $\sqrt s=192$~GeV.
We adopt the SM parameters from Ref.~\cite{PDG}.
We assume the fourth-generation quarks $T$ and $B$ of the HP model to be
degenerate with mass $m_T=m_B=250$~GeV.
The other input parameters are varied.
We present the cross sections and their radiative corrections in the $G_F$
formulation of the on-shell renormalization scheme; {\it i.e.}, we eliminate
$\alpha$ via the relation
\begin{equation}
G_F=\frac{\pi\alpha}{\sqrt2s_w^2M_W^2}\,\frac{1}{1-\Delta r},
\end{equation}
where $\Delta r$ \cite{sir} contains those radiative corrections to the muon
lifetime which the SM or its extensions introduce on top of the purely
photonic corrections from within the Fermi model.
Specifically, we fix $\alpha=\sqrt2G_Fs_w^2M_W^2/\pi$ and, in turn, substitute
$\delta e/e\to\delta e/e-\Delta r/2$ in Eq.~(\ref{CT}).
In Ref.~\cite{Hansi}, $\Delta r$ has been calculated together with the $S$,
$T$, and $U$ parameters \cite{STU} in the HP model.
Reference~\cite{Hansi} also provides general expressions for these quantities 
in terms of the vacuum polarizations $\Pi_T^{V_1V_2}$ defined in
Eq.~(\ref{PiT}), which allow us to obtain the corresponding results for the MR
model.

We start by restating the tree-level cross sections of processes~(i)--(iv).
Figure 1(a) shows the differential cross sections
${\rm d}\sigma/{\rm d}\cos\theta$ as functions of the scattering angle
$\theta$ in the c.m.\ frame.
The results for $\gamma\gamma$ and $\gamma Z$ production exhibit $t$- and 
$u$-channel poles in the forward and backward directions, at $\theta=0^\circ$
and $180^\circ$, which are artifacts of our approximation of neglecting the
electron mass.
However, these poles are avoided if we impose the experimental acceptance cut
$|\cos\theta|<0.966$, which excludes the regions around the beam pipe not
covered by the detectors.
The integrated cross sections of processes~(i)--(iv), evaluated with this
angular cut, are displayed in Fig.~1(b) as functions of $\sqrt s$.
In the LEP2 energy range and above, $e^+e^-\to W^+W^-$ has the largest cross
section of all the vector-boson pair-production processes.
This process is being extensively studied at LEP2, since it offers a unique
opportunity to experimentally establish the non-Abelian nature of the SM,
through the $\gamma WW$ and $ZWW$ vertices.
In the remainder of this section, we thus focus our attention on
$e^+e^-\to W^+W^-$.

The parameter spaces of the HP and MR models are significantly constrained by
a wealth of electroweak high-precision data.
It is convenient to extract these constraints by means of the $S$, $T$, and $U$
parameters \cite{STU}.
According to a recent global data analysis \cite{Paul}, at the 68\% confidence 
level, the new-physics contributions to $S$, $T$, and $U$ are confined within 
the ranges
\begin{equation}
\label{STU}
-0.43 < S_{\rm new} < 0.14,\qquad
-0.38 < T_{\rm new} < 0.28 ,\qquad
-0.36 < U_{\rm new} < 0.48 .
\end{equation}
These limits include the uncertainty in the Higgs-boson mass, which is varied
in the range $60 < M_H < 1000$~GeV.
In addition, there exist limits from direct searches for new heavy fermions at
LEP1 and LEP2.
Specifically, we require the masses of all fourth-generation neutrinos to be
larger than $M_Z/2$, so as to avoid an observable contribution to the
invisible width of the $Z$ boson.
Furthermore, we take the masses of the fourth-generation charged leptons to be
larger than $\sqrt s/2$, for otherwise they would have been produced and
directly detected at LEP2.
Finally, we must exclude the production of two heavy neutrinos, $n_i$ and 
$n_j$, by $e^+e^-$ annihilation via a virtual $Z$ boson followed by the decay
of $n_i$ and/or $n_j$ into a heavy charged lepton $E$ and an off-shell $W$ 
boson.
In summary,
\begin{itemize}
\item[(i)] $m_{n_i}>M_Z/2$ for all $n_i$;
\item[(ii)] $m_E>\sqrt s/2$;
\item[(iii)] if $m_{n_i}+m_{n_j}<\sqrt s$, then $m_{n_i},m_{n_j}<m_E$.
\end{itemize}

The shaded areas in Figs.~2 and 3 indicate the allowed regions of parameter
space for the HP and MR models, respectively.
In the four parts of Fig.~2, the $(m_E,m_D)$ plane is scanned for
$m_M=0$, 0.25, 1, and 5~TeV, respectively.
The case $m_M=0$ corresponds to a fourth-generation Dirac neutrino.
The effects of the various constraints mentioned above are easily recognized.
The $S$ parameter determines the left edges of the allowed areas,
while the right edges are controlled by the $T$ parameter.
Finally, constraints from direct searches confine the shaded regions from
below by horizontal lines.
Note that the $U$ parameter does not yield any further constraint on the HP
model.

Next, we present a similar analysis for the MR model, varying $m_1$ and $m_2$
continuously and $m_M$ and $m_E$ discretely.
In Figs.~3(a)--(d), the $(m_1,m_2)$ plane is scanned for
$m_M=0$, 0.25, 1, and 5~TeV, respectively.
The four parts of each figure refer to $m_E=0.1$, 0.25, 0.5, and 1~TeV, 
respectively.
The reflection symmetry with respect to the diagonal $m_1=m_2$, which is
common to all figures, is a property of the neutrino mass matrix of the MR
model.
Similarly to the HP model, the $U$ parameter does not lead to any actual
constraint.
Obviously, the allowed parameter space of the MR model is larger than that of
the HP model.
This is mainly due to the facts that, in the MR model, fourth-generation
quarks are absent and that, besides the $T$ parameter, also the $S$ parameter
may take negative values.

In the remainder of this section, we study the radiative corrections to the
cross section of $e^+e^-\to W^+W^-$ induced by the heavy fermions of the HP
and MR models taking into account the above constraints.
Figures~4 and 5 display the angular dependence of the radiative corrections to
the differential cross section for the HP and MR models, respectively, with
typical parameter sets in compliance with the above bounds.
We see that the angular dependence is asymmetric.
In general, the loop effects are more significant in the backward direction.
In the HP model, they may be as large as 1\%, for $m_E=500$~GeV,
$m_D=250$~GeV, and $m_M=5$~TeV, while, in the MR model, they may not exceed
the 0.1\% level, for $m_E=m_1=m_2=250$~GeV and $m_M=0$.

In Figs.~6(a) and (b), contours of constant correction to the integrated cross
section are drawn in the $(m_E,m_D)$ plane of the HP model with $m_M=0$ and
1~TeV, respectively.
As in Fig.~2, the shaded areas indicate the allowed parameter space.
Figures~7(a)--(c) describe a similar analysis for the $(m_1,m_2)$ plane of
the MR model with
(a) $m_E=250$~GeV and $m_M=0$,
(b) $m_E=250$~GeV and $m_M=1$~TeV, and
(c) $m_E=1$~TeV and $m_M=10$~TeV, respectively.
From this analysis, we conclude that the shifts in the integrated cross
section of $e^+e^-\to W^+W^-$ due to the fourth fermion generations of the
constrained HP and MR models cannot exceed 0.1\%.

\section{Conclusions\label{sec:concl}}

We have studied quantum effects due to heavy exotic leptons in the reactions
$e^+e^-\to \gamma\gamma$, $\gamma Z$, $ZZ$, and $W^+W^-$ at LEP2 energies
within two viable scenarios of new physics, the HP and MR models.
Both models extend the lepton sector of the SM by additional sequential
isodoublets.
These models also admit the presence of Majorana mass terms of the order of
the electroweak scale, which may lead to lepton-number violating signals
through the production of heavy Majorana neutrinos at TeV energies.
If the direct production of heavy neutrinos predicted by the HP and MR models
is impossible at present energies, such neutrinos may still give rise to
significant shifts in the oblique electroweak parameters $S$, $T$, and $U$.
Exploiting recent experimental information on the $S$, $T$, and $U$
parameters, we have systematically constrained the parameter spaces of the HP
and MR models.
We have then quantitatively analyzed the loop-induced shifts in the cross
section of $e^+e^- \to  W^+W^-$ which arise in the HP and MR models thus 
constrained.
We have found that these shifts cannot exceed the benchmark of 0.1\%.
In conclusion, cross-section measurements at LEP2 are unlikely to improve the
bounds on the parameters of the HP and MR models already established by
electroweak high-precision data.
In other words, observable deviations from the SM predictions would require
alternative explanations.
This could not be anticipated without explicit calculation because, in
contrast to $Z$-resonance and low-energy physics, vector-boson pair production
at LEP2 is already sensitive to the triple-gauge-boson couplings at the tree
level.

\bigskip

\noindent  
{\bf Acknowledgements.}
We would like to thank Probir Roy for useful discussions and Ansgar Denner for
providing a computer code which allowed us to check our numerical results for
$e^+e^-\to W^+W^-$ in the SM limit.

\newpage 

\def\theequation{\Alph{section}.\arabic{equation}}
\begin{appendix}
\setcounter{equation}{0}
\section{Mass eigenvalues and eigenvectors in the MR model\label{eigenvals}}

The couplings $g^\pm_{n_in_jZ}$ and $g^\pm_{En_iW^-}$ of the MR model given in
Eq.~(\ref{gparMR}) implicitly depend on the masses of the heavy neutrinos
$n_i$ and the charged lepton $E$, through the unitary matrix $U$.
Therefore, it is important to express all the parameters, including the mixing
angles, in terms of a minimal set of independent variables.
We choose this set to be $m_M$, $m_1$, $m_2$, and $m_E$, which we assume to be
real so as to avoid CP violation.

Defining the auxiliary variables
\begin{eqnarray}
p&=&\frac{1}{3}\left(\frac{m_M^2}{3}+m_1^2+m_2^2+m_E^2\right),
\nonumber\\
q&=&\frac{m_M}{3}\left(\frac{m_M^2}{9}+\frac{m_1^2+m_2^2}{2}-m_E^2\right)
-m_1m_2m_E,
\nonumber\\
\phi&=&\arccos\frac{q}{p^{3/2}},
\end{eqnarray}
we may write the three eigenvalues $m_{n_k}$ ($k=1,2,3$) of the mass matrix
in Eq.~(\ref{LmassMR}) in the compact form
\begin{equation}
m_{n_k}=\frac{m_M}{3}+2\sqrt p\cos\frac{\phi+2\pi(k-1)}{3}.
\end{equation}
Solving the eigenvalue problem for the MR neutrino mass matrix, we obtain the
following set of real, orthogonal vectors:
\begin{equation}
\label{Ok}
O_k={\cal N}_k\left(m_1m_{n_k}-m_2m_E,m_2m_{n_k}-m_1m_E,m^2_{n_k}-m_E^2\right),
\end{equation}
where ${\cal N}_k$ are normalization constants defined such that $O_kO^T_k=1$.
In fact, $O_k$ are eigenvectors of the neutrino mass matrix, with real
eigenvalues, $\pm m_{n_k}$.
If we assume that $m_M$, $m_1$, $m_2$, and $m_E$ are all positive, then we
find that the physical mass eigenvectors are $O_1$, $iO_2$, and $O_3$.
In this parameter range, the $3\times 3$ unitary matrix $U$ is given by
\begin{equation}
\label{Umatrix}
U=(O_1^T,iO_2^T,O_3^T).
\end{equation}

\setcounter{equation}{0}
\section{Standard matrix elements\label{stdmatrix}}

On general grounds, the radiatively corrected helicity amplitudes
${\cal M}^\kappa(\lambda_1,\lambda_2,s,t)$ of processes~(i)--(iv) can all be
written as linear combinations of ten independent standard matrix elements.
Following Ref.~\cite{Sack}, we define
\begin{eqnarray} 
\label{StandardME} 
{\cal M}_0^\kappa&=&
\bar{v}(p_+){\not\!\varepsilon}_1({\not\!k}_1-{\not\!p}_+)
{\not\!\varepsilon}_2{\rm P}_\kappa u(p_-),
\nonumber\\
{\cal M}_1^\kappa&=& 
\bar{v}(p_+){\not\!k}_1{\rm P}_\kappa u(p_-)\,
\varepsilon_1\cdot\varepsilon_2,
\nonumber\\
{\cal M}_2^\kappa&=&
\bar{v}(p_+){\not\!\varepsilon}_1{\rm P}_\kappa u(p_-)\,
\varepsilon_2\cdot k_1,
\nonumber\\
{\cal M}_3^\kappa&=&
-\bar{v}(p_+){\not\!\varepsilon}_2{\rm P}_\kappa u(p_-)\,
\varepsilon_1\cdot k_2,
\nonumber\\
{\cal M}_4^\kappa&=&
\bar{v}(p_+){\not\!\varepsilon}_1{\rm P}_\kappa u(p_-)\,
\varepsilon_2\cdot p_-,
\nonumber\\
{\cal M}_5^\kappa&=&
-\bar{v}(p_+){\not\!\varepsilon}_2{\rm P}_\kappa u(p_-)\,
\varepsilon_1\cdot p_+,
\nonumber\\
{\cal M}_6^\kappa&=& 
\bar{v}(p_+){\not\!k}_1{\rm P}_\kappa u(p_-)\,\varepsilon_1\cdot p_+\,
\varepsilon_2\cdot p_-,
\nonumber\\
{\cal M}_7^\kappa&=& 
\bar{v}(p_+){\not\!k}_1{\rm P}_\kappa u(p_-)\,\varepsilon_1\cdot p_+\,
\varepsilon_2\cdot k_1,
\nonumber\\
{\cal M}_8^\kappa&=& 
\bar{v}(p_+){\not\!k}_1{\rm P}_\kappa u(p_-)\,\varepsilon_1\cdot k_2\,
\varepsilon_2\cdot p_-,
\nonumber\\
{\cal M}_9^\kappa&=& 
\bar{v}(p_+){\not\!k}_1{\rm P}_\kappa u(p_-)\,\varepsilon_1\cdot k_2\,
\varepsilon_2\cdot k_1,
\end{eqnarray}
where we have suppressed the arguments of $\varepsilon_1(k_1,\lambda_1)$ and
$\varepsilon_2(k_2,\lambda_2)$.
In our application, ${\cal M}_6^\kappa$, ${\cal M}_7^\kappa$, and
${\cal M}_8^\kappa$ do not occur.
This would be subject to change if we also allowed for inter-family mixing
between standard and exotic fermions.

\setcounter{equation}{0}
\section{One-loop functions\label{oneloop}}

In this paper, we evaluate the loop amplitudes using dimensional
regularization in $D$ space-time dimensions along with the reduction algorithm
of Ref.~\cite{PV}.
In contrast to Ref.~\cite{PV}, we use the Minkowskian metric,
$g^{\mu\nu}={\rm diag}(1,-1,\ldots,-1)$.
As usual, we introduce an unphysical 't~Hooft mass scale, $\mu$, to keep the
coupling constants dimensionless.

The vector two-point function occurring in Eq.~(\ref{PiT}) is defined as
\cite{KP}
\begin{eqnarray}
\Pi_V(q^2,m_1,m_2)&=&
\frac{1}{12\pi^2}\left\{\left[q^2-\frac{m_1^2+m_2^2}{2}
+3m_1m_2-\frac{\left(m_1^2-m_2^2\right)^2}{2q^2}\right]B_0(q,m_1,m_2)\right.
\nonumber\\
&&+m_1^2\left(-1+\frac{m_1^2-m_2^2}{2q^2}\right)B_0(0,m_1,m_1)
+m_2^2\left(-1+\frac{m_2^2-m_1^2}{2q^2}\right)
\nonumber\\
&&\left.\times
B_0(0,m_2,m_2)-\frac{q^2}{3}+\frac{\left(m_1^2-m_2^2\right)^2}{2q^2}\right\},
\label{eqpiv}
\end{eqnarray}
where
\begin{equation}
B_0(p,m_1,m_2)=\frac{(2\pi\mu)^{4-D}}{i\pi^2}\int{\rm d}^Dq\,
\frac{1}{\left(q^2-m_1^2\right)\left[(q+p)^2-m_2^2\right]}
\end{equation}
is the standard two-point scalar integral \cite{PV}.

Next, we list analytic expressions for the triple-gauge-boson vertex functions
${\cal V}_{\kappa_1\kappa_2\kappa_3}^\kappa$ of Eq.~(\ref{Vkappa}) in terms of
standard three-point integrals \cite{PV}.
There is a total of 16 possible helicity combinations.
Keeping $\kappa=\pm$ generic, we have
\begin{eqnarray}
{\cal V}^\kappa_{\pm\pm\pm}&=&
2\left[\frac{2}{3}-2\left(2C_2^0+C_3^{01}+C_3^{02}\right)
-M_1^2\left(C_1^1+2C_2^1+C_3^1+C_3^{12}\right)
\right.\nonumber\\
&&\left.
-M_2^2\left(C_1^2+2C_2^2+C_3^2+C_3^{21}\right)
+\left(s-M_1^2-M_2^2\right)\left(2C_2^{12}+C_3^{12}+C_3^{21}\right)
\vphantom{\frac{2}{3}}\right]{\cal M}^\kappa_1
\nonumber\\
&&+2\left[-\frac{2}{3}+2\left(C_2^0-C_3^{01}\right)
-M_1^2\left(C_2^1+C_3^1\right)+M_2^2\left(C_1^2+C_2^2-C_3^{21}\right)
\right.\nonumber\\
&&\left.
+\left(s-M_1^2-M_2^2\right)C_3^{12}
\vphantom{\frac{2}{3}}\right]{\cal M}^\kappa_2
\nonumber\\
&&+2\left[-\frac{2}{3}+2\left(C_2^0-C_3^{02}\right)
+M_1^2\left(C_1^1+C_2^1-C_3^{12}\right)-M_2^2\left(C_2^2+C_3^2\right)
\right.\nonumber\\ 
&&\left.
+\left(s-M_1^2-M_2^2\right)C_3^{21}
\vphantom{\frac{2}{3}}\right]{\cal M}^\kappa_3
-8\left(C_2^{12}+C_3^{12}+C_3^{21}\right){\cal M}^\kappa_9
\nonumber\\
&&\pm2\kappa\left[-\frac{1}{3}+2\left(C_2^0 + 3C_3^{01}\right)
+M_1^2\left(C_2^1+C_3^1\right)+M_2^2\left(C_1^2+C_2^2+2C_2^{12}+C_3^{21}\right)
\right.\nonumber\\
&&\left.
-\left(s-M_1^2-M_2^2\right)C_3^{12}
\vphantom{\frac{1}{3}}\right]
\left({\cal M}^\kappa_0+{\cal M}^\kappa_1-{\cal M}^\kappa_2
-2{\cal M}^\kappa_5\right)
\nonumber\\
&&\pm2\kappa\left[-\frac{1}{3}+2\left(C_2^0+3C_3^{02}\right)
+M_1^2\left(C_1^1+C_2^1+2C_2^{12}+C_3^{12}\right)+M_2^2\left(C_2^2+C_3^2\right)
\right.\nonumber\\
&&\left.
-\left(s-M_1^2-M_2^2\right)C_3^{21}
\vphantom{\frac{1}{3}}\right]
\left({\cal M}^\kappa_0+{\cal M}^\kappa_1-{\cal M}^\kappa_3
-2{\cal M}^\kappa_4\right),\nonumber\\
{\cal V}^\kappa_{\pm\mp\mp}&=&
2m_im_j\left[\left(C_0+C_1^1+C_1^2\right){\cal M}^\kappa_1
+C^1_1{\cal M}^\kappa_2 -\left(C_0+C_1^2\right){\cal M}^\kappa_3
\pm\kappa C_1^1
\right.\nonumber\\
&&\left.\times
\left({\cal M}^\kappa_0+{\cal M}^\kappa_1-{\cal M}^\kappa_2
-2{\cal M}^\kappa_5\right)
\pm\kappa\left(C_0+C_1^2\right)
\left({\cal M}^\kappa_0+{\cal M}^\kappa_1-{\cal M}^\kappa_3
-2{\cal M}^\kappa_4\right)\right],\nonumber\\
{\cal V}^\kappa_{\mp\mp\pm}&=&
2m_jm_k\left[-\left(C_1^1+C_1^2\right){\cal M}^\kappa_1
+C^1_1{\cal M}^\kappa_2+C_1^2{\cal M}^\kappa_3\pm\kappa C_1^1
\right.\nonumber\\
&&\left.\times
\left({\cal M}^\kappa_0+{\cal M}^\kappa_1-{\cal M}^\kappa_2
-2{\cal M}^\kappa_5\right)
\pm\kappa C_1^2\left({\cal M}^\kappa_0+{\cal M}^\kappa_1-{\cal M}^\kappa_3
-2{\cal M}^\kappa_4\right)\right],\nonumber\\
{\cal V}^\kappa_{\mp\pm\mp}&=&
2m_km_i\left[\left(C_0+C_1^1+C_1^2\right){\cal M}^\kappa_1
-\left(C_0+C^1_1\right){\cal M}^\kappa_2+C_1^2{\cal M}^\kappa_3
\pm\kappa\left(C_0+C_1^1\right)
\right.\nonumber\\
&&\left.
\times\left({\cal M}^\kappa_0+{\cal M}^\kappa_1-{\cal M}^\kappa_2
-2{\cal M}^\kappa_5\right)\pm\kappa C_1^2
\left({\cal M}^\kappa_0+{\cal M}^\kappa_1-{\cal M}^\kappa_3-
{\cal M}^\kappa_4\right)\right] .
\label{eqv}
\end{eqnarray}
Here, ${\cal V}_{\pm\pm\pm}^\kappa$ stands for ${\cal V}_{+++}^\kappa$ or
${\cal V}_{---}^\kappa$ and similarly for the other expressions.
The $C$ functions appearing in Eq.~(\ref{eqv}) are the scalar coefficients in
the Lorentz decompositions \cite{PV} of the standard three-point tensor
integrals
\begin{eqnarray}
\label{C0}
\lefteqn{\{C_0,C_\mu,C_{\mu\nu},C_{\mu\nu\rho}\}(k_1,-k_2,m_i,m_j,m_k)}
\nonumber\\
&=&\frac{(2\pi\mu)^{4-D}}{i\pi^2}\int{\rm d}^Dq\,
\frac{\{1,q_\mu,q_\mu q_\nu,q_\mu q_\nu q_\rho\}}
{\left(q^2-m_i^2\right)\left[(q+k_1)^2-m_j^2\right]
\left[(q-k_2)^2-m_k^2\right]}.
\end{eqnarray}
In the notation of Ref.~\cite{Sack}, we have
\begin{eqnarray}
C_\mu&=&k_{1\mu}C_1^1-k_{2\mu}C_1^2,
\nonumber\\
C_{\mu\nu}&=&
g_{\mu\nu}C_2^0+k_{1\mu}k_{1\nu}C_2^1+k_{2\mu}k_{2\nu}C_2^2
-(k_{1\mu}k_{2\nu}+k_{2\mu}k_{1\nu})C_2^{12},
\nonumber\\
C_{\mu\nu\rho}&=&
(g_{\mu\nu}k_{1\rho}+g_{\nu\rho}k_{1\mu}+g_{\mu\rho}k_{1\nu})C_3^{01}
-(g_{\mu\nu}k_{2\rho}+g_{\nu\rho}k_{2\mu}+g_{\mu\rho}k_{2\nu})C_3^{02}
\nonumber\\
&&+k_{1\mu}k_{1\nu}k_{1\rho}C_3^1-k_{2\mu}k_{2\nu}k_{2\rho}C_3^2
-(k_{1\mu}k_{1\nu}k_{2\rho}+k_{1\mu}k_{2\nu}k_{1\rho}
+k_{2\mu}k_{1\nu}k_{1\rho})C_3^{12}
\nonumber\\
&&+(k_{2\mu}k_{2\nu}k_{1\rho}+k_{2\mu}k_{1\nu}k_{2\rho}
+k_{1\mu}k_{2\nu}k_{2\rho})C_3^{21}.
\end{eqnarray}
Analytic expressions for the Lorentz coefficients may be found in
Ref.~\cite{Sack}.

\setcounter{equation}{0}
\section{Renormalization constants\label{renorm}}

We work in the on-shell renormalization scheme, which uses the fine-structure
constant $\alpha=e^2/(4\pi)$ and the physical particle masses as basic
parameters.
As usual, we express the bare parameters and wave functions, which carry the 
subscript 0, in terms of the respective renormalized quantities and
counterterms as
\begin{eqnarray}
e_0&=&e+\delta e,\qquad
s_{w,0}=s_w+\delta s_w,\qquad
g_{eeZ,0}^\kappa=g_{eeZ}^\kappa+\delta g_{eeZ}^\kappa,
\nonumber\\
M_{W,0}^2&=&M_W^2+\delta M_W^2,\qquad
M_{Z,0}^2=M_Z^2+\delta M_Z^2,\qquad
W_0^\pm=W^\pm\left(1+\frac{\delta Z_W}{2}\right),
\nonumber\\
\left(
\begin{array}{c}
Z_0\\
A_0
\end{array}
\right)
&=&
\left(
\begin{array}{cc}
1+\delta Z_{ZZ}/2&\delta Z_{Z\gamma}/2\\
\delta Z_{\gamma Z}/2&1+\delta Z_{\gamma\gamma}/2
\end{array}
\right)
\left(
\begin{array}{c}
Z\\
A
\end{array}
\right).
\label{eqct}
\end{eqnarray}
In the absence of mixing between the standard and exotic fermion generations,
the electron mass and wave function are not affected by the new physics at one 
loop.
The counterterms in Eq.~(\ref{eqct}) can all be expressed in terms of the 
transverse vacuum-polarization functions defined in Eq.~(\ref{PiT}), as
\begin{eqnarray}
\delta M_W^2&=&{\rm Re}\Pi_T^{WW}\left(M_W^2\right),\qquad
\delta M_Z^2={\rm Re}\Pi_T^{ZZ}\left(M_Z^2\right),
\nonumber\\
\delta Z_W&=&\left.-{\rm Re}\frac{\partial\Pi_T^{WW}(p^2)}{\partial p^2}
\right|_{p^2=M_W^2},\qquad
\delta Z_{ZZ}=\left.-{\rm Re}\frac{\partial\Pi_T^{ZZ}(p^2)}{\partial p^2}
\right|_{p^2=M_Z^2},
\nonumber\\
\delta Z_{\gamma\gamma}&=&
\left.-\frac{\partial\Pi_T^{\gamma\gamma}(p^2)}{\partial p^2}\right|_{p^2=0},
\qquad
\delta Z_{Z\gamma}=2\frac{\Pi_T^{\gamma Z}(0)}{M_Z^2},\qquad
\delta Z_{\gamma Z}=-2{\rm Re}\frac{\Pi_T^{\gamma Z}(M_Z^2)}{M_Z^2},
\nonumber\\
\frac{\delta e}{e}&=&-\frac{1}{2}
\left(\delta Z_{\gamma\gamma}+\frac{s_w}{c_w}\delta Z_{Z\gamma}\right),\qquad
\frac{\delta s_w}{s_w}=\frac{c_w^2}{2s_w^2}\left(\frac{\delta M_Z^2}{M_Z^2}
-\frac{\delta M_W^2}{M_W^2}\right),
\nonumber\\
\frac{\delta g_{eeZ}^+}{g_{eeZ}^+}&=&\frac{1}{c_w^2}\,\frac{\delta s_w}{s_w},
\qquad
\frac{\delta g_{eeZ}^-}{g_{eeZ}^-}=\frac{1}{c_w^2\left(s_w^2-c_w^2\right)}\,
\frac{\delta s_w}{s_w}.
\label{eqos}
\end{eqnarray}

\end{appendix}

\newpage

\centerline{\Large{\bf Figure Captions}}
\vspace{-0.2cm}
\newcounter{fig}
\begin{list}{\rm {\bf Fig.~\arabic{fig}:}}{\usecounter{fig}
\labelwidth1.6cm \leftmargin2.5cm \labelsep0.4cm \itemsep0ex plus0.2ex }

\item {\bf(a)} Differential and {\bf(b)} integrated cross sections of
$e^+e^-\to\gamma\gamma$, $\gamma Z$, $ZZ$, and $W^+W^-$ in the Born 
approximation.

\item Allowed parameter space of the HP model with $m_M=0$, 0.25, 1, and 
5~TeV.

\item Allowed parameter space of the MR model with $m_E=0.1$, 0.25, 0.5, and
1~TeV and {\bf(a)} $m_M=0$, {\bf(b)} $m_M=0.25$~TeV, {\bf(c)} $m_M=1$~TeV,
and {\bf(d)} $m_M=5$~TeV.

\item Radiative correction relative to the differential cross section of
$e^+e^-\to W^+W^-$ in the HP model as a function of the scattering angle.

\item Radiative correction relative to the differential cross section of
$e^+e^-\to W^+W^-$ in the MR model as a function of the scattering angle.

\item Contour levels of the radiative correction relative to the integrated
cross section of $e^+e^-\to W^+W^-$ in the HP model with
{\bf(a)} $m_M=0$ and
{\bf(b)} $m_M=1$~TeV.
The allowed regions of parameter space are shaded.

\item Contour levels of the radiative correction relative to the integrated
cross section of $e^+e^-\to W^+W^-$ in the RM model with
{\bf(a)} $m_E=0.25$~TeV and $m_M=0$,
{\bf(b)} $m_E=0.25$~TeV and $m_M=1$~TeV, and
{\bf(c)} $m_E=1$~TeV and $m_M=10$~TeV, respectively.
The allowed regions of parameter space are shaded.
\end{list}

\newpage

\begin{figure}[p]
  \begin{center}
    \leavevmode
    \epsfig{file=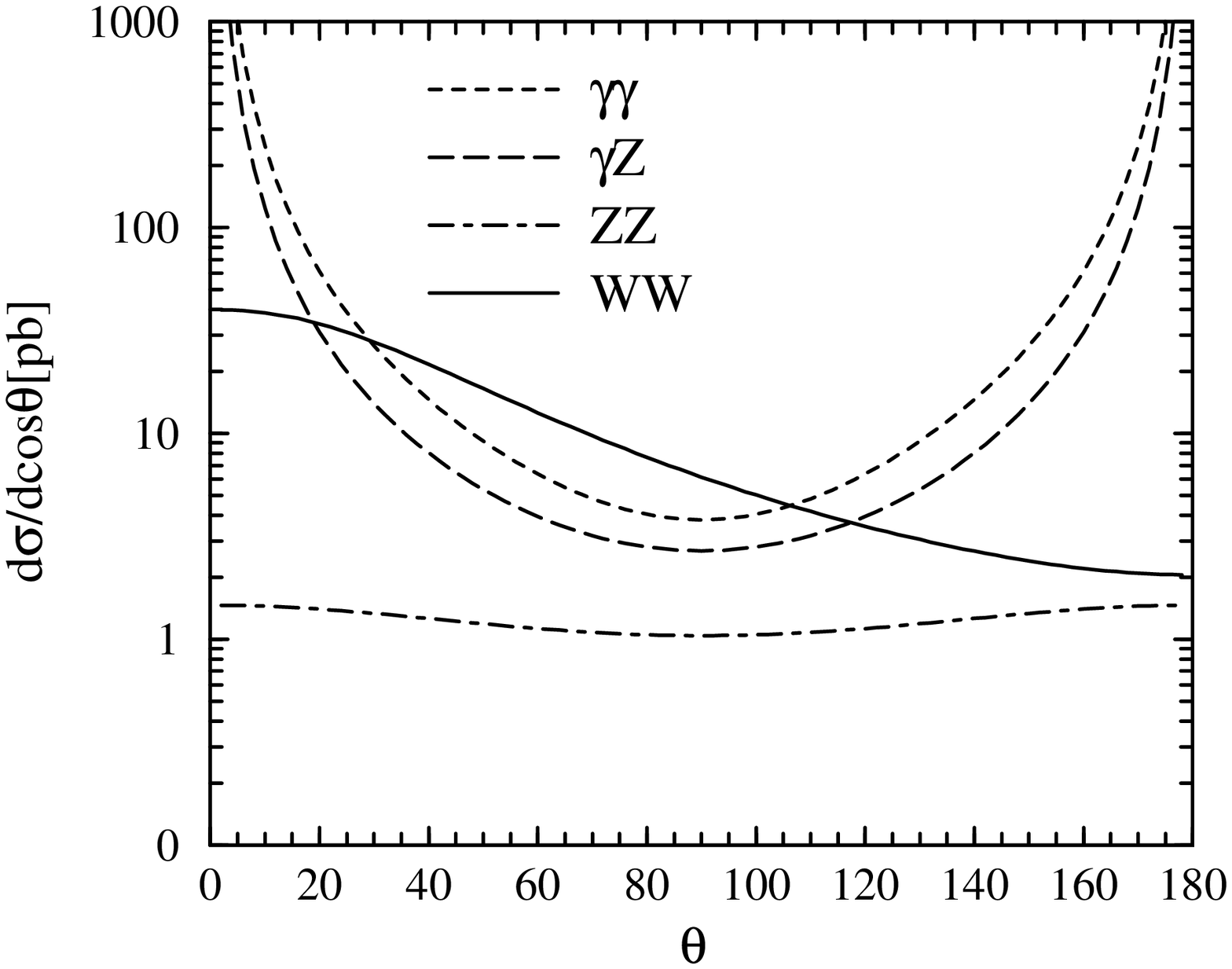,width=9.5cm,height=9.5cm}
    \centerline{\Large\bf Fig.~1a}
%
    \epsfig{file=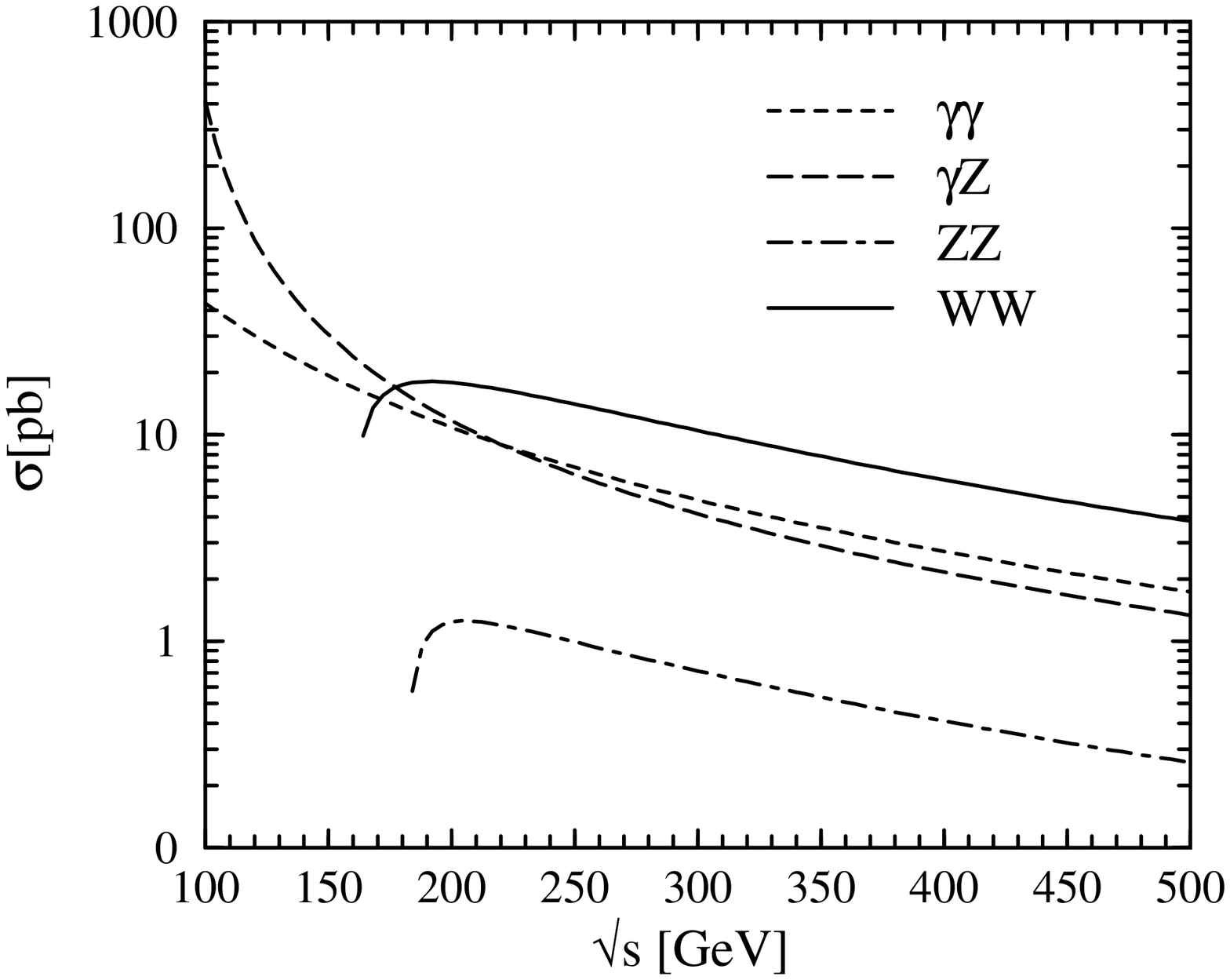,width=9.5cm,height=9.5cm}
    \centerline{\Large\bf Fig.~1b}
  \end{center}
\end{figure}

\begin{figure}[p]
  \begin{center}
    \begin{tabular}{cc}
      \parbox{8cm}{
        \epsfig{file=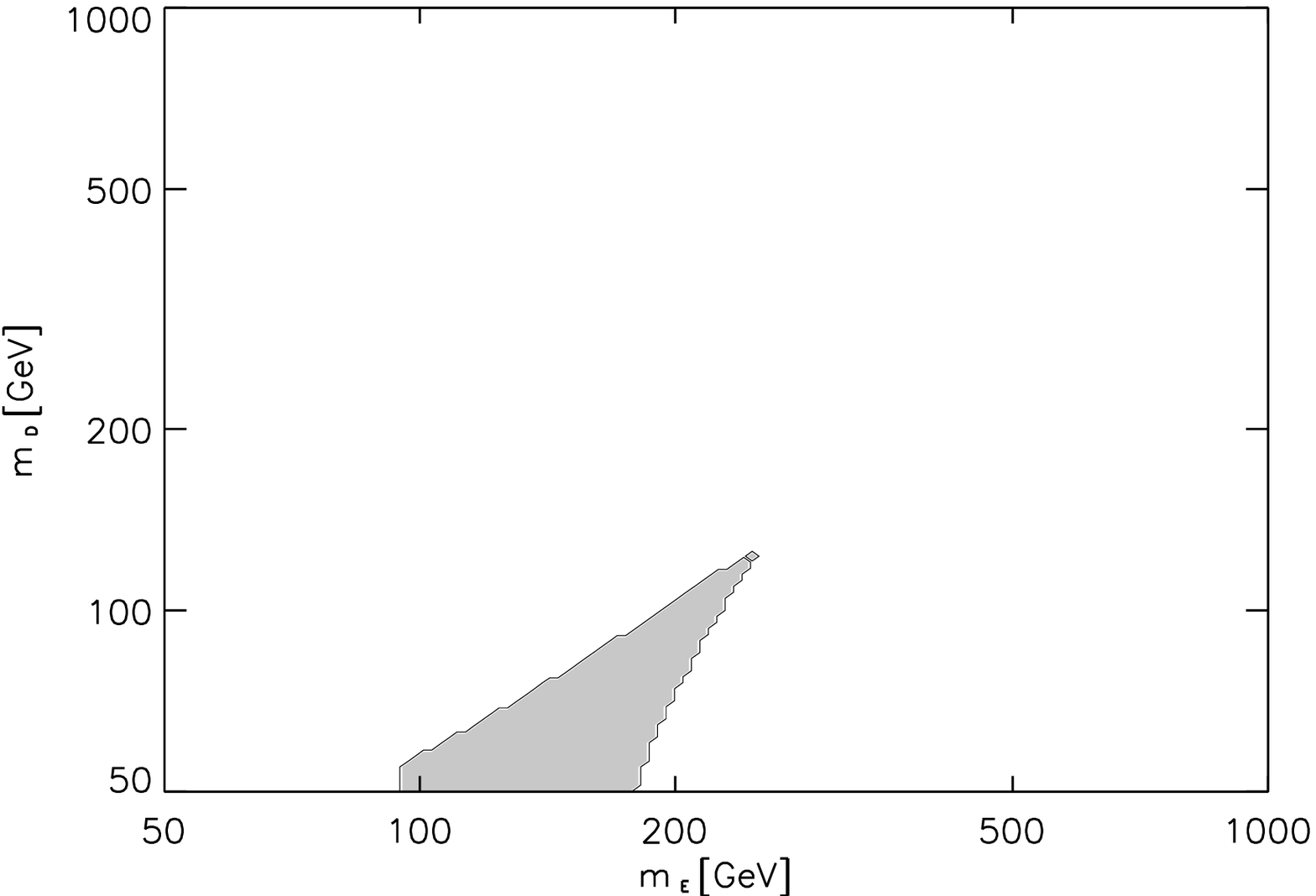,width=7.5cm,height=7.5cm}
        }
      &
      \parbox{8cm}{
        \epsfig{file=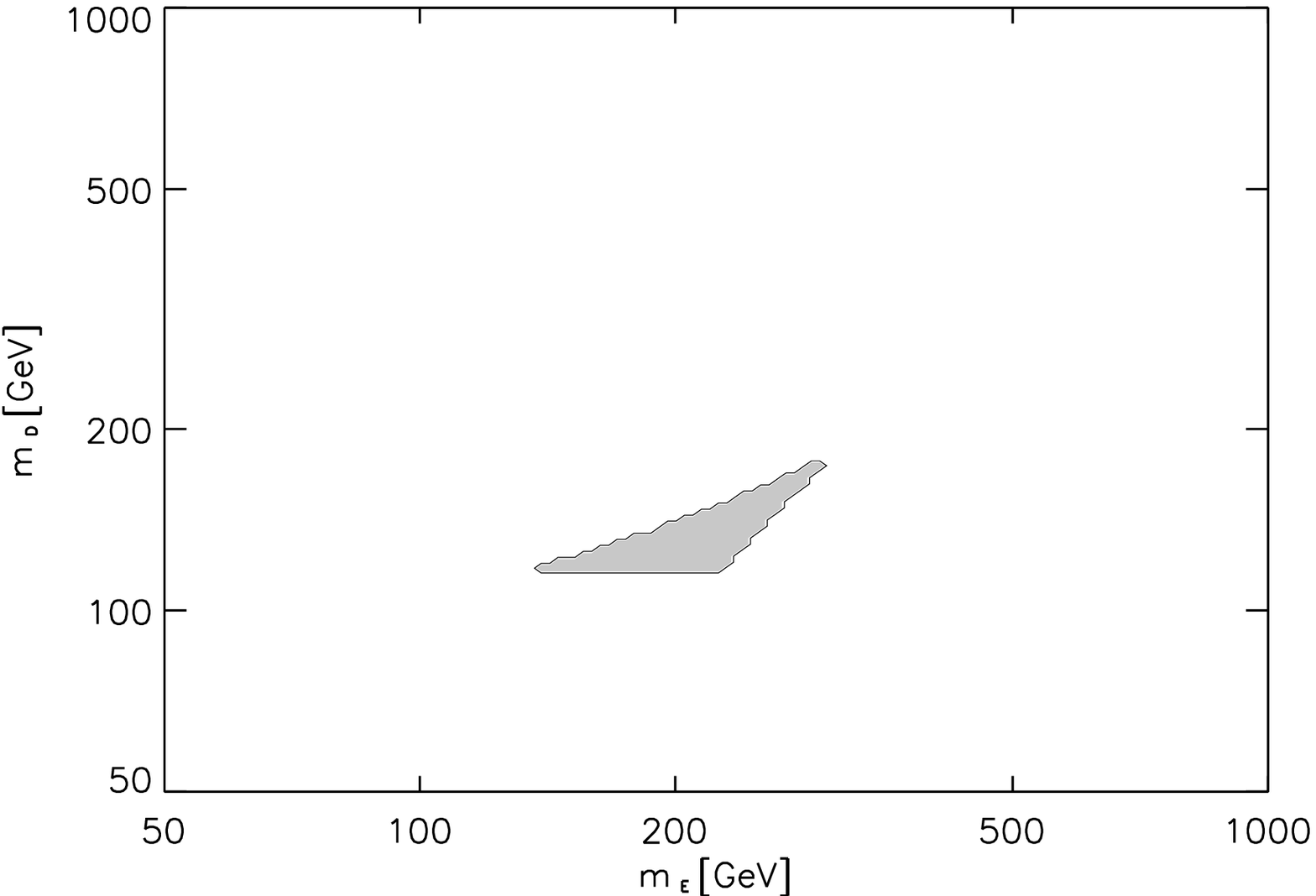,width=7.5cm,height=7.5cm}
        }
      \\
      \parbox{8cm}{
        \epsfig{file=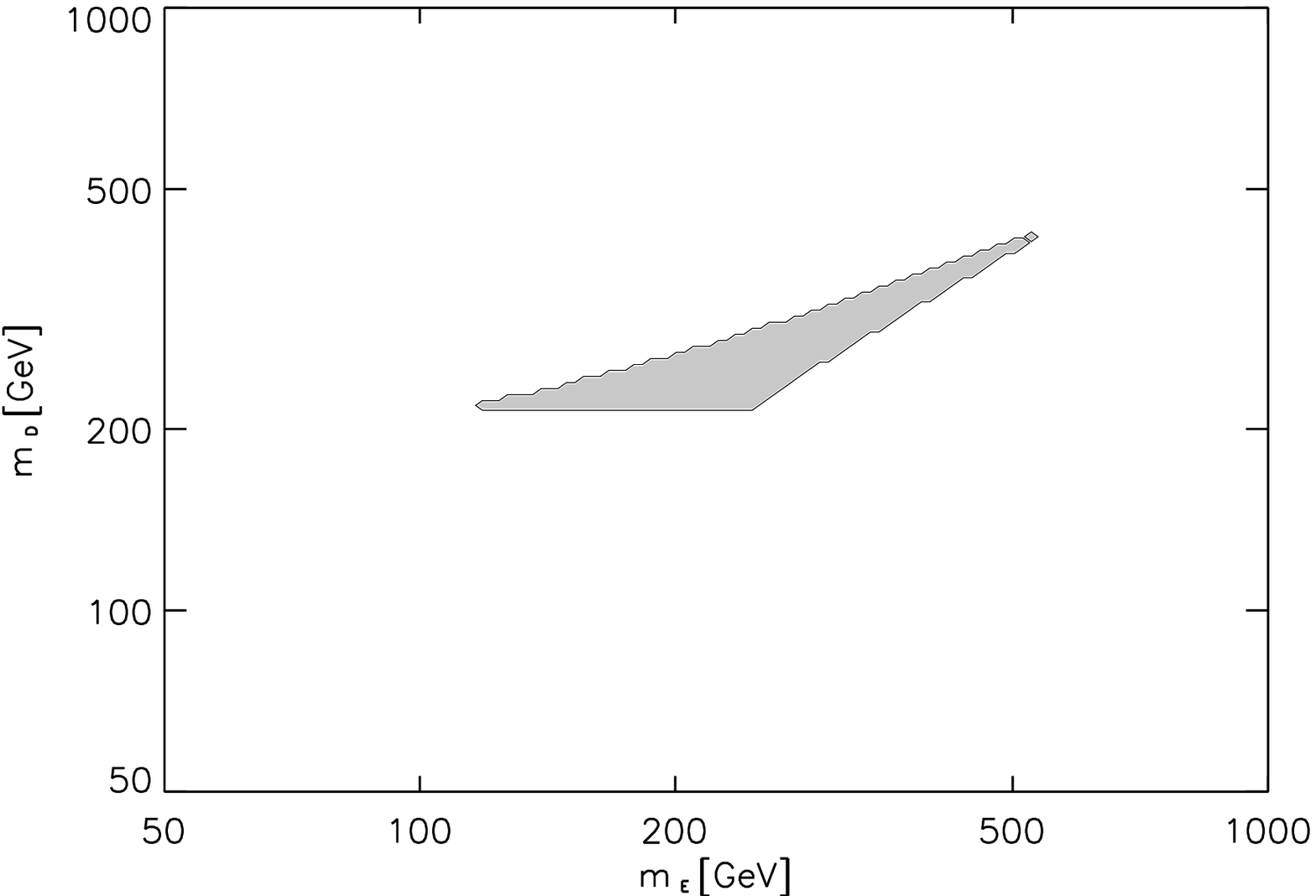,width=7.5cm,height=7.5cm}
        }
      &
      \parbox{8cm}{
        \epsfig{file=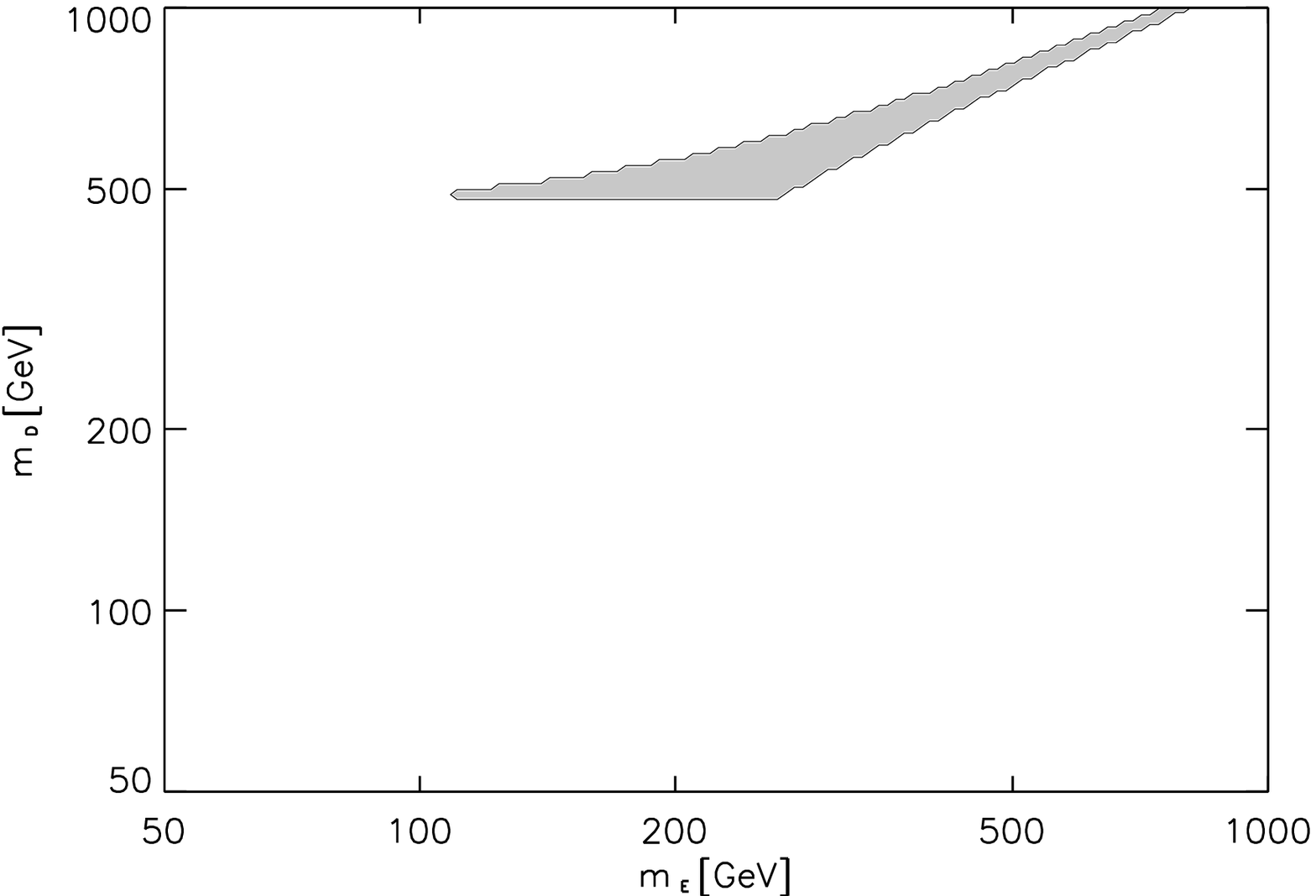,width=7.5cm,height=7.5cm}
        }
    \end{tabular}
   \end{center}
\bigskip
\centerline{\Large\bf Fig.~2}
\end{figure}

\begin{figure}[p]
  \begin{center}
    \begin{tabular}{cc}
      \parbox{8cm}{
        \epsfig{file=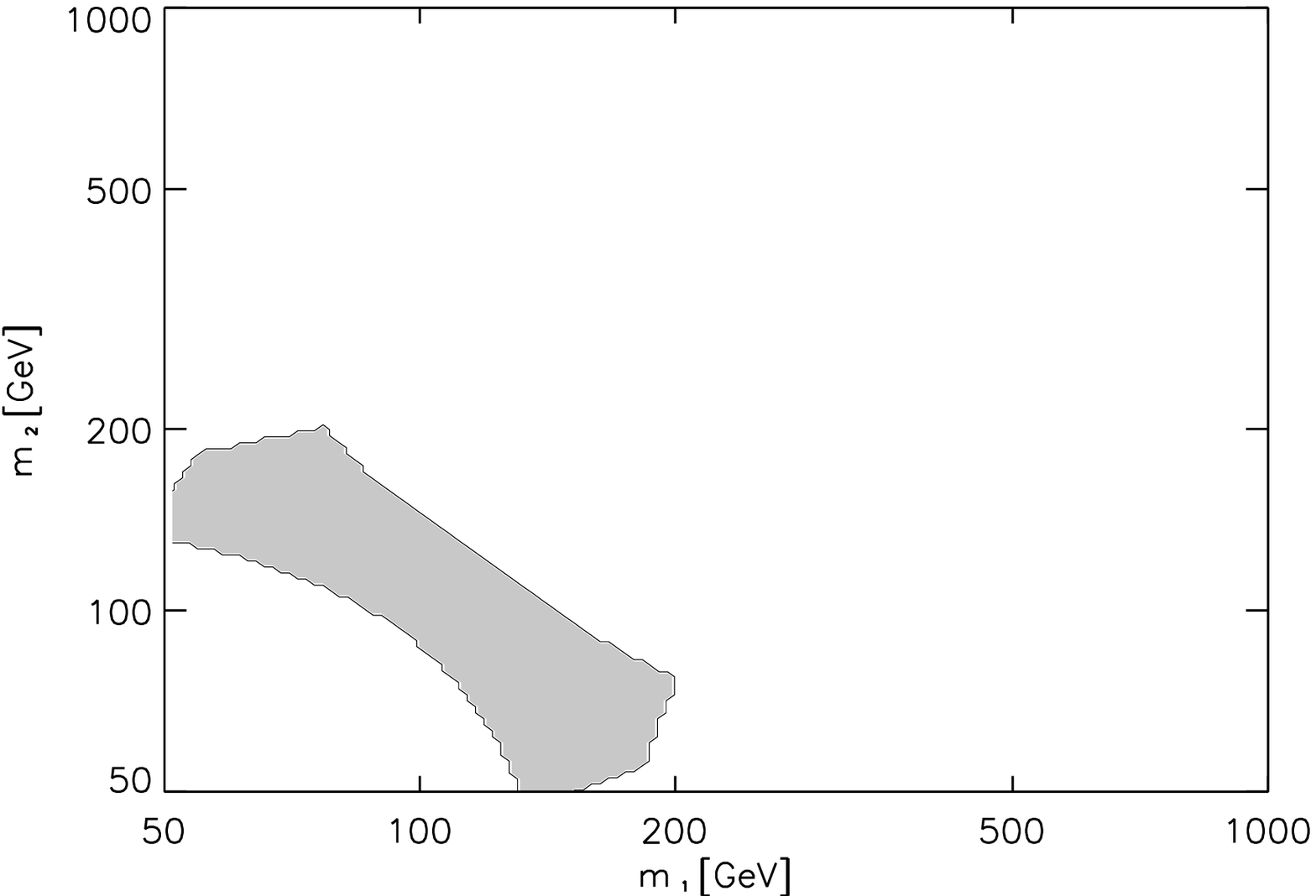,width=7.5cm,height=7.5cm}
        }
      &
      \parbox{8cm}{
        \epsfig{file=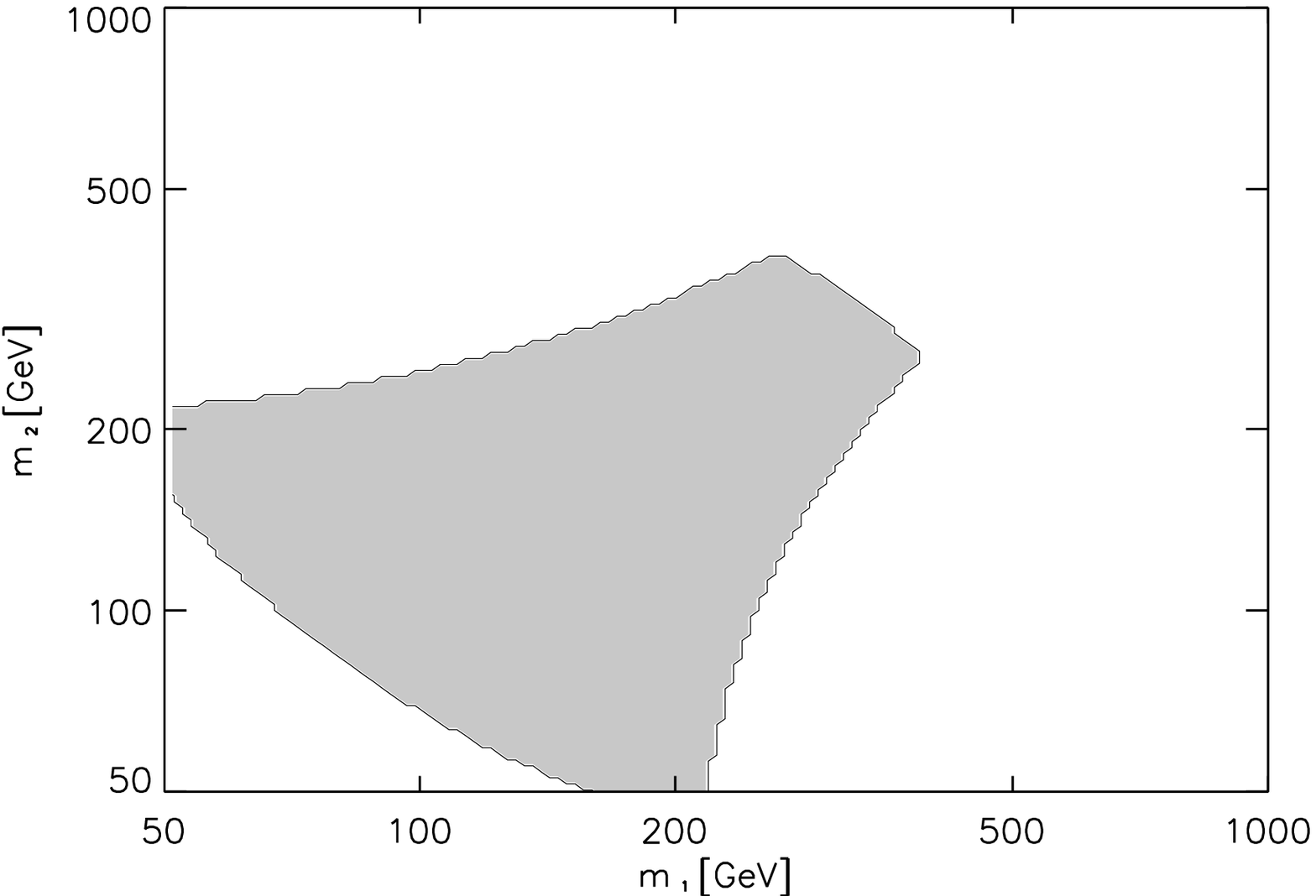,width=7.5cm,height=7.5cm}
        }
      \\
      \parbox{8cm}{
        \epsfig{file=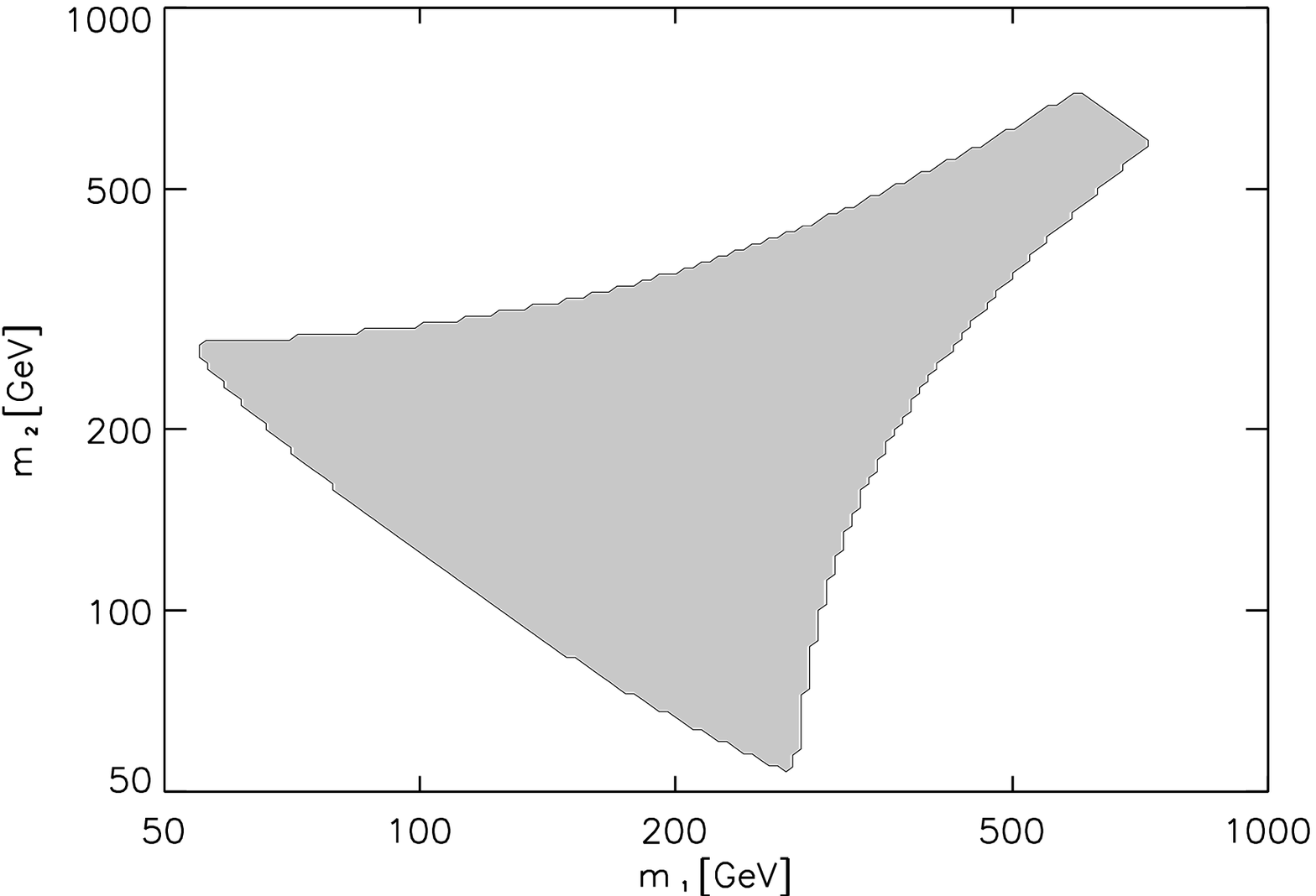,width=7.5cm,height=7.5cm}
        }
      &
      \parbox{8cm}{
        \epsfig{file=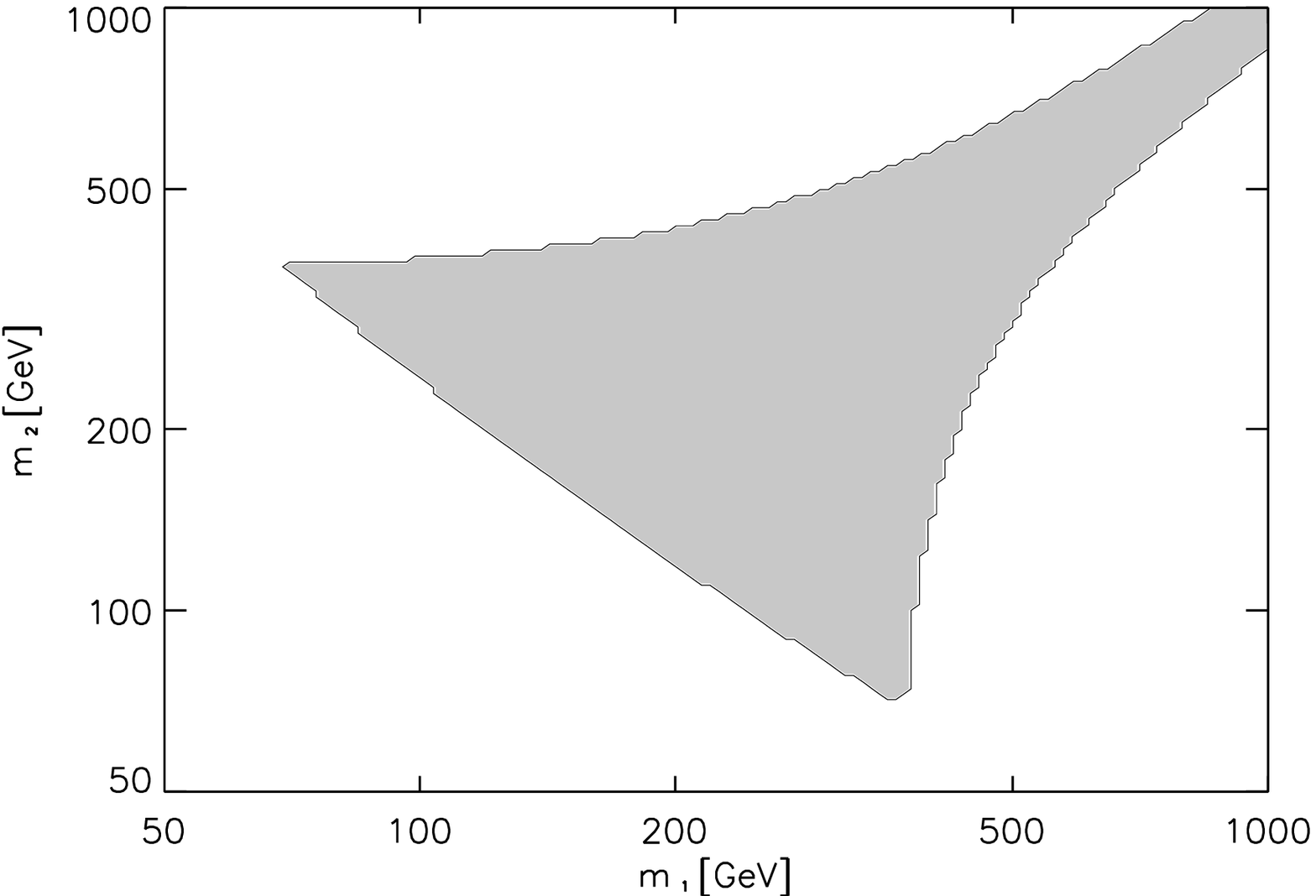,width=7.5cm,height=7.5cm}
        }
    \end{tabular}
  \end{center}
\bigskip
\centerline{\Large\bf Fig.~3a}
\end{figure}

\begin{figure}[p]
  \begin{center}
    \begin{tabular}{cc}
      \parbox{8cm}{
        \epsfig{file=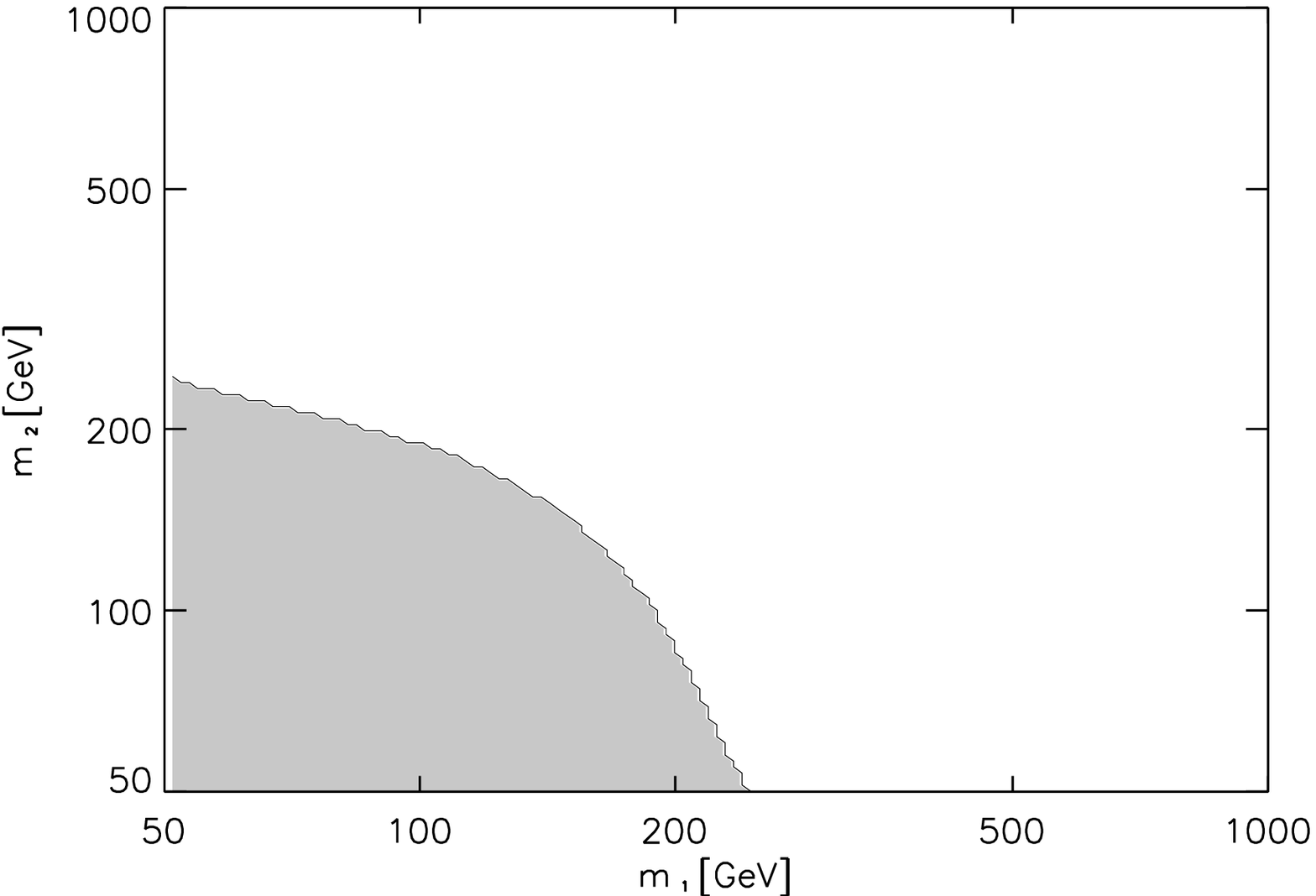,width=7.5cm,height=7.5cm}
        }
      &
      \parbox{8cm}{
        \epsfig{file=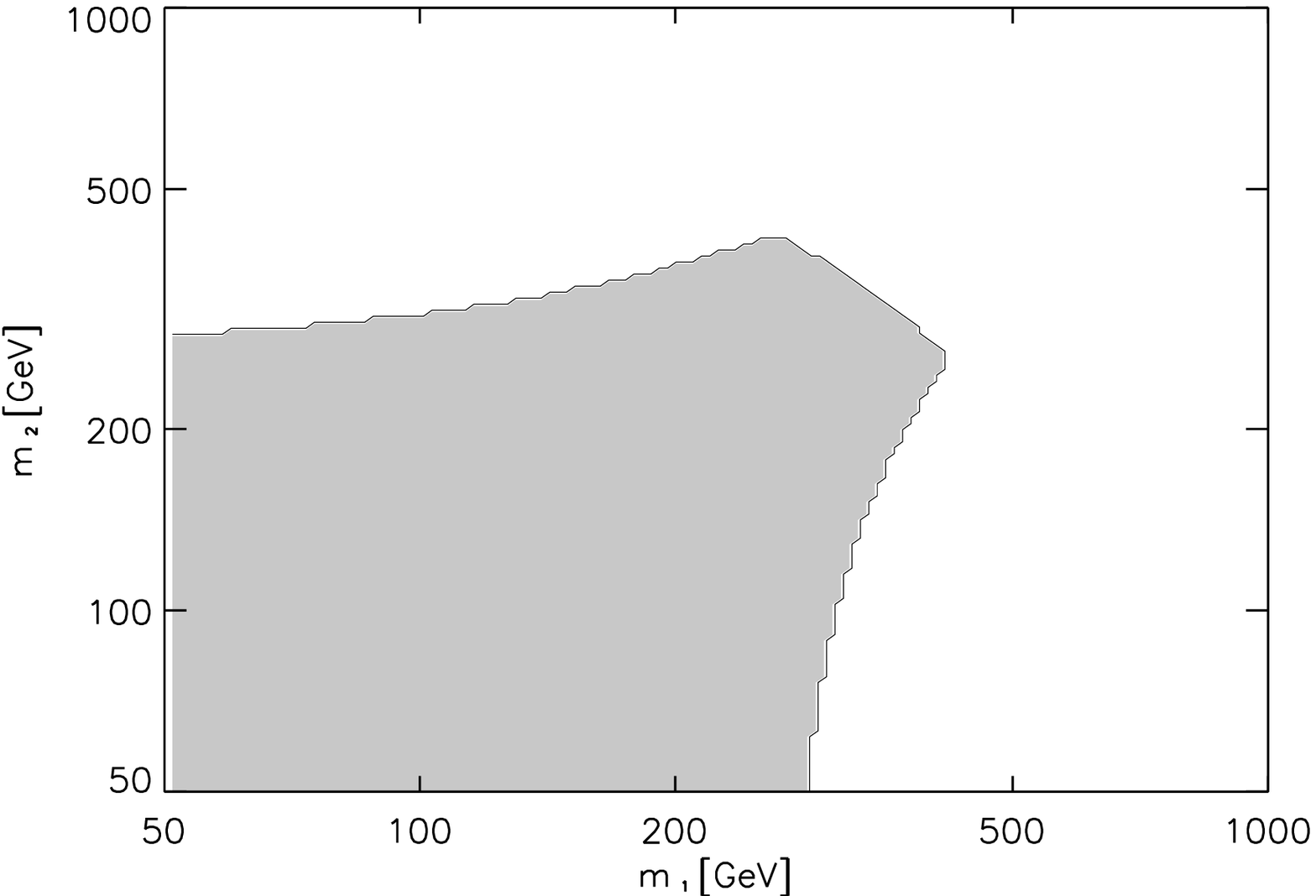,width=7.5cm,height=7.5cm}
        }
      \\
      \parbox{8cm}{
        \epsfig{file=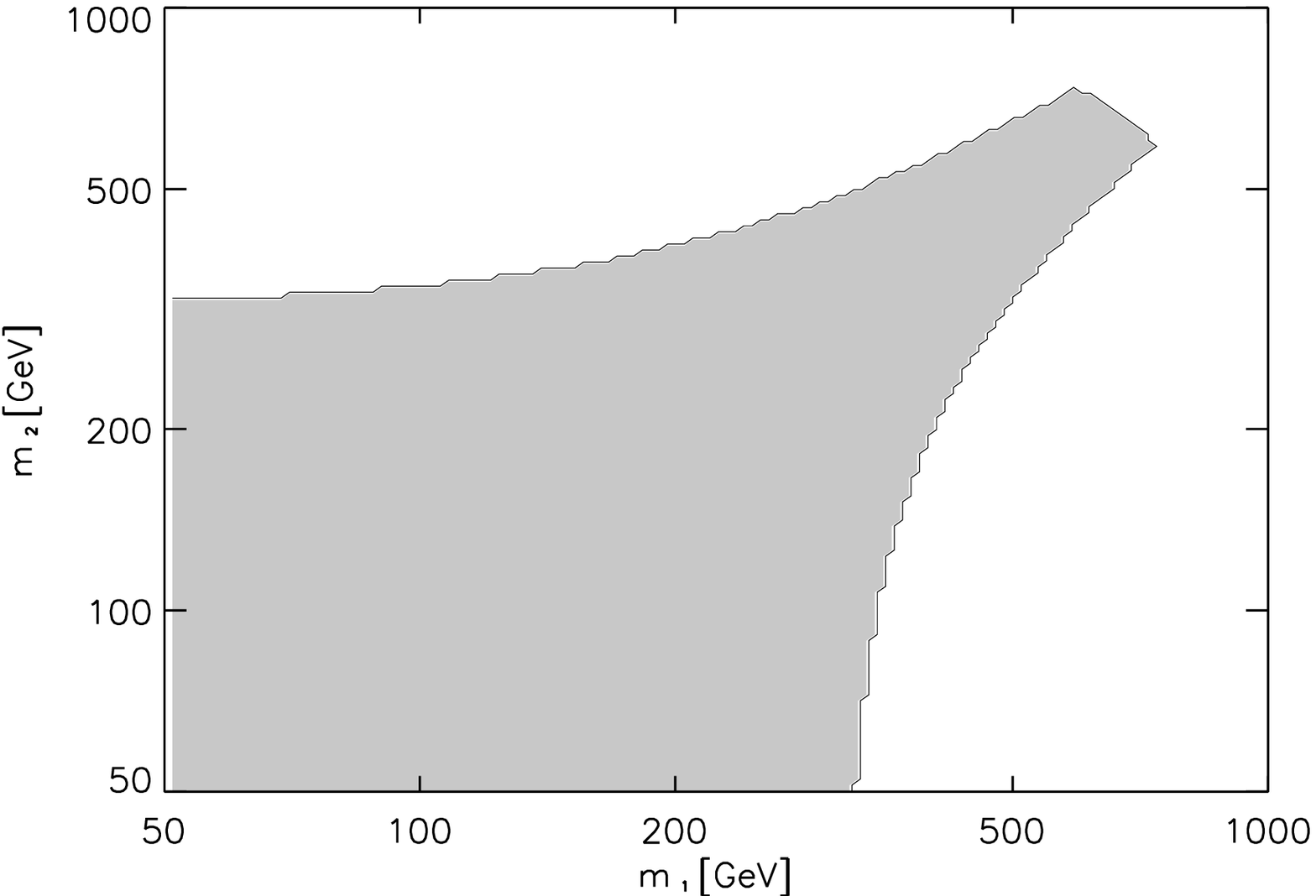,width=7.5cm,height=7.5cm}
        }
      &
      \parbox{8cm}{
        \epsfig{file=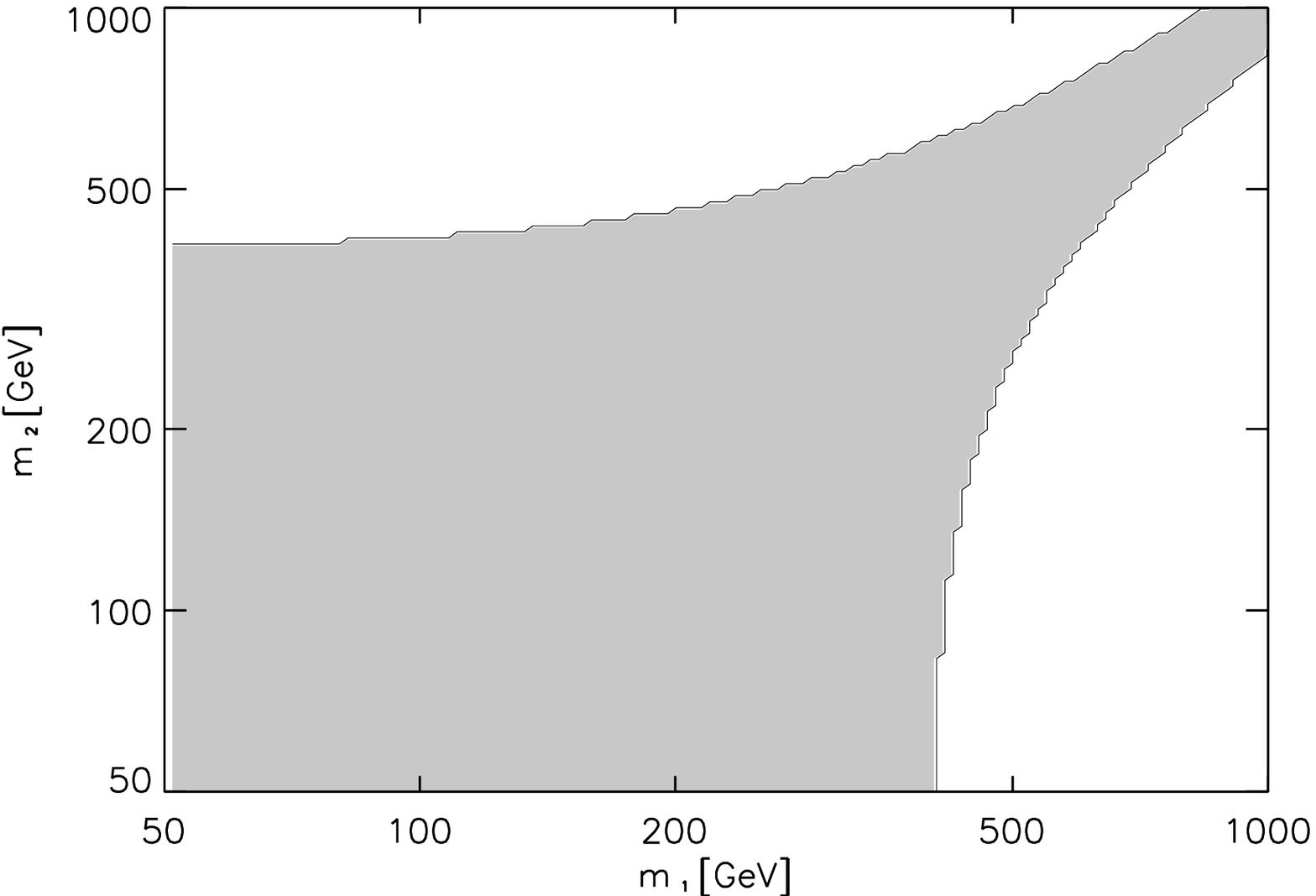,width=7.5cm,height=7.5cm}
        }
    \end{tabular}
  \end{center}
\bigskip
    \centerline{\Large\bf Fig.~3b}
\end{figure}

\begin{figure}[p]
  \begin{center}
    \begin{tabular}{cc}
      \parbox{8cm}{
        \epsfig{file=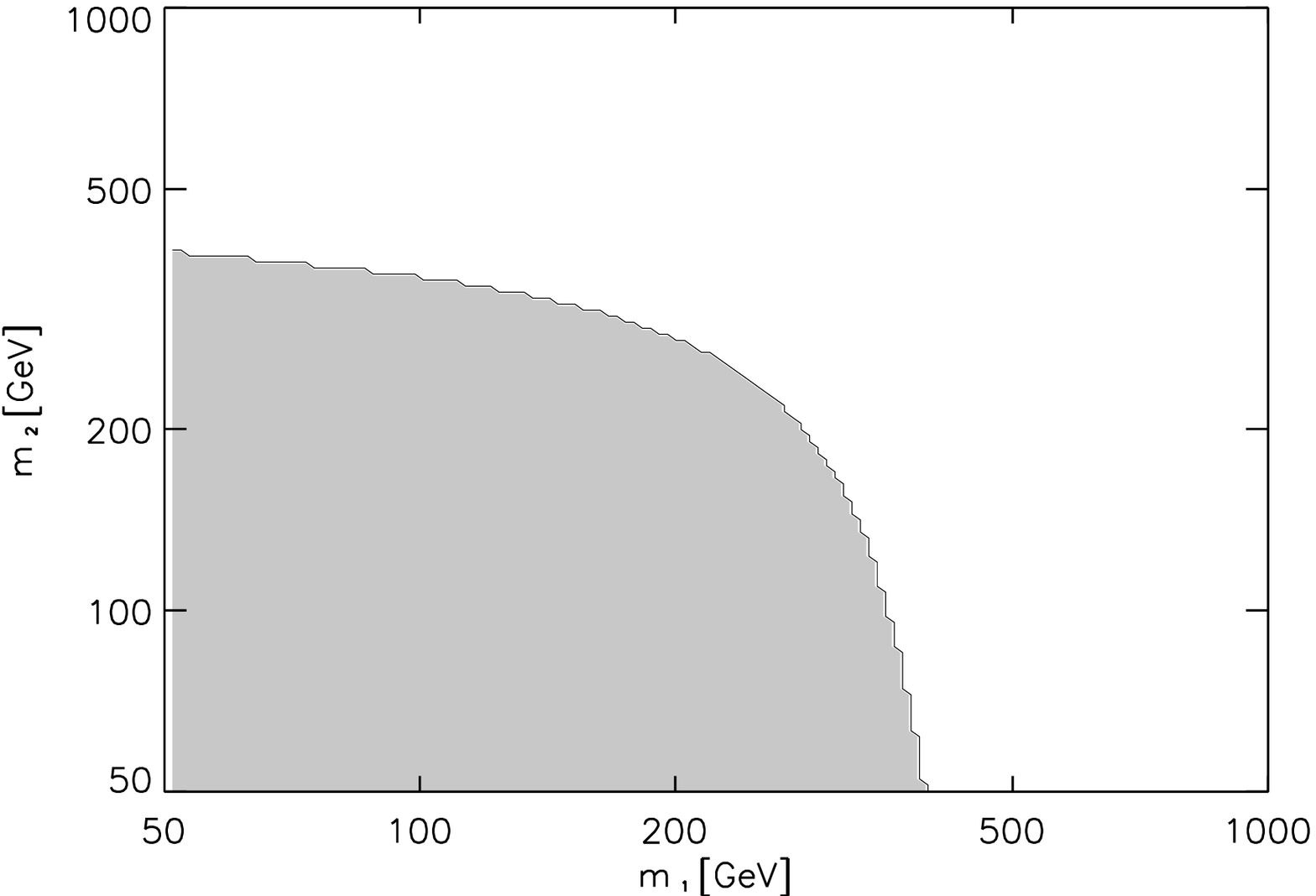,width=7.5cm,height=7.5cm}
        }
      &
      \parbox{8cm}{
        \epsfig{file=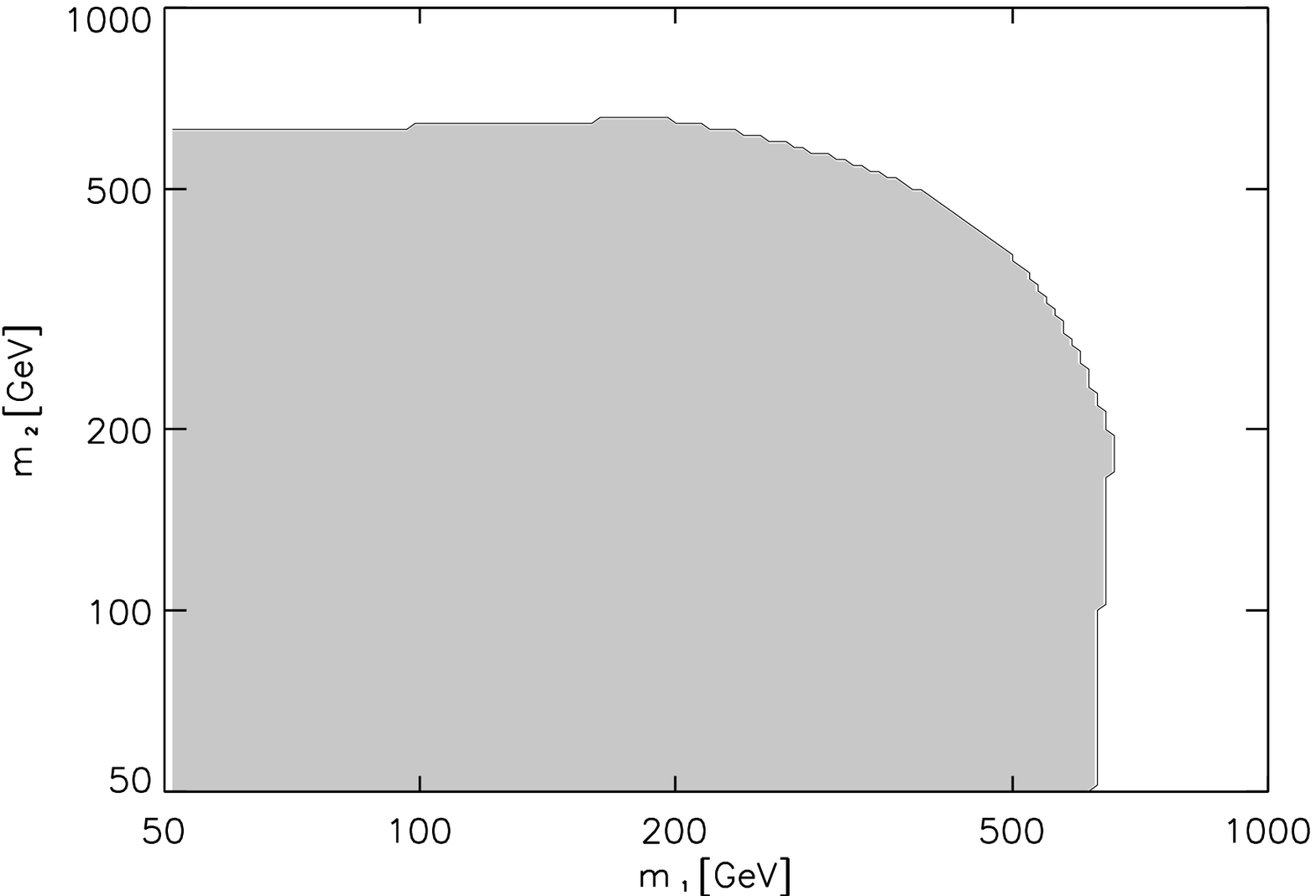,width=7.5cm,height=7.5cm}
        }
      \\
      \parbox{8cm}{
        \epsfig{file=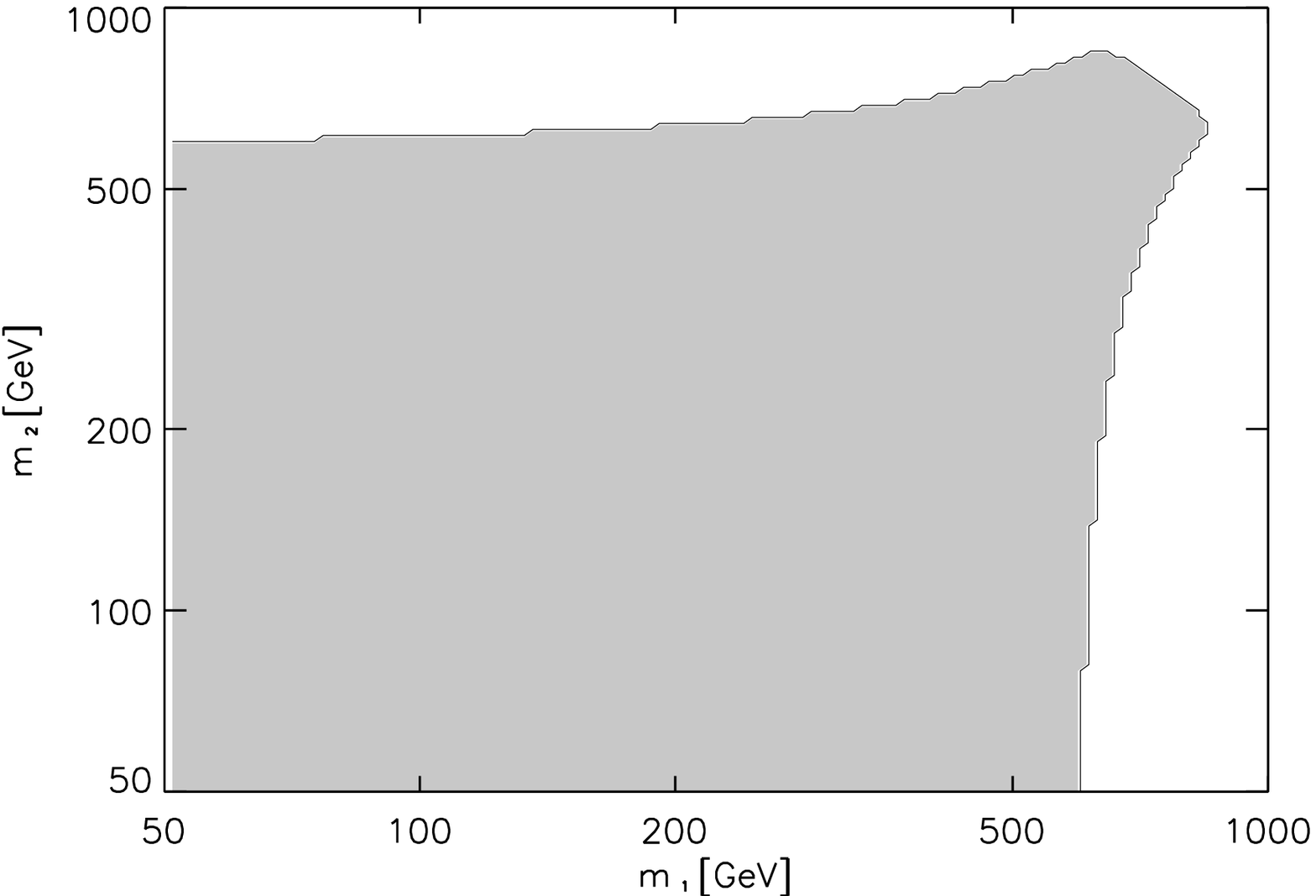,width=7.5cm,height=7.5cm}
        }
      &
      \parbox{8cm}{
        \epsfig{file=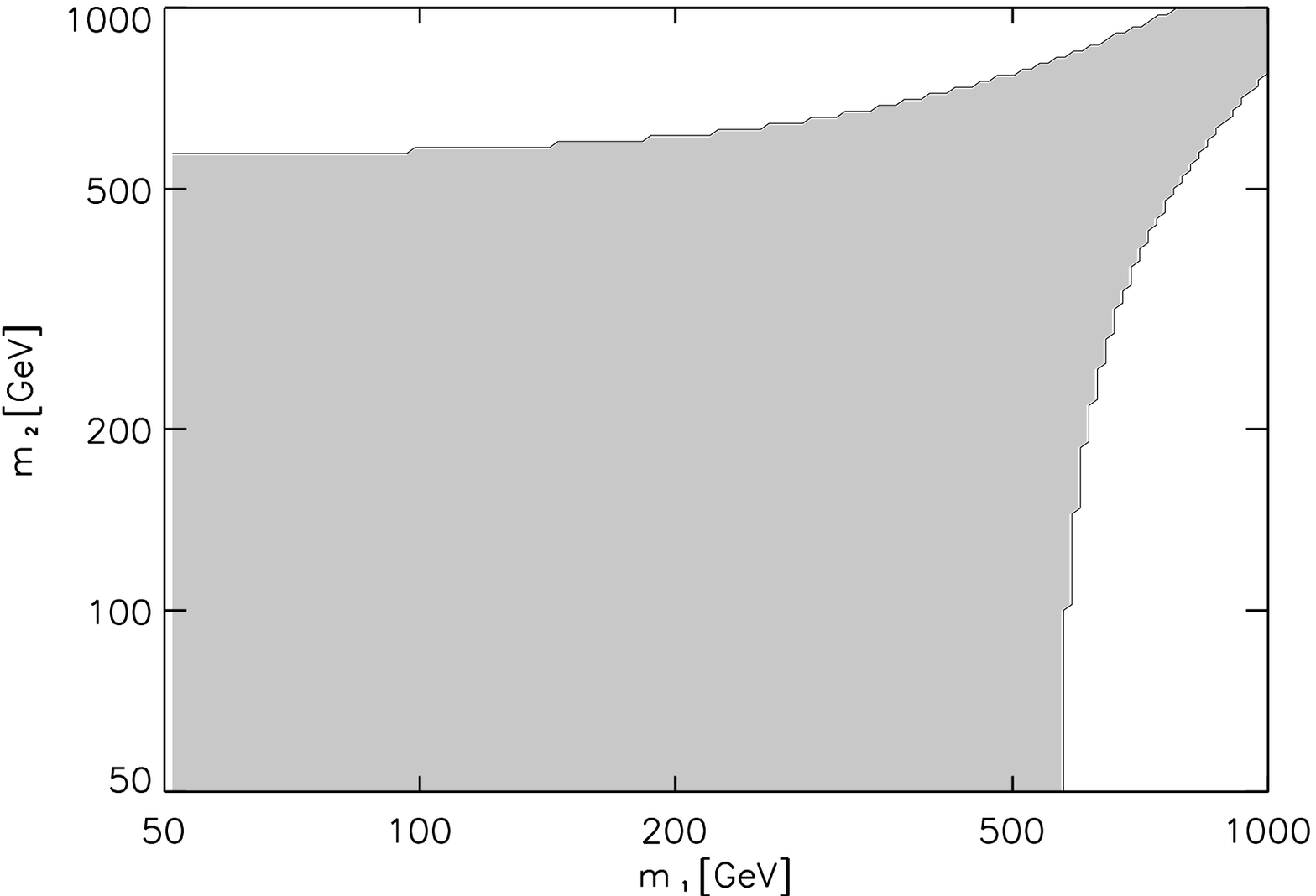,width=7.5cm,height=7.5cm}
        }
    \end{tabular}
  \end{center}
\bigskip
    \centerline{\Large\bf Fig.~3c}
\end{figure}

\begin{figure}[p]
  \begin{center}
    \begin{tabular}{cc}
      \parbox{8cm}{
        \epsfig{file=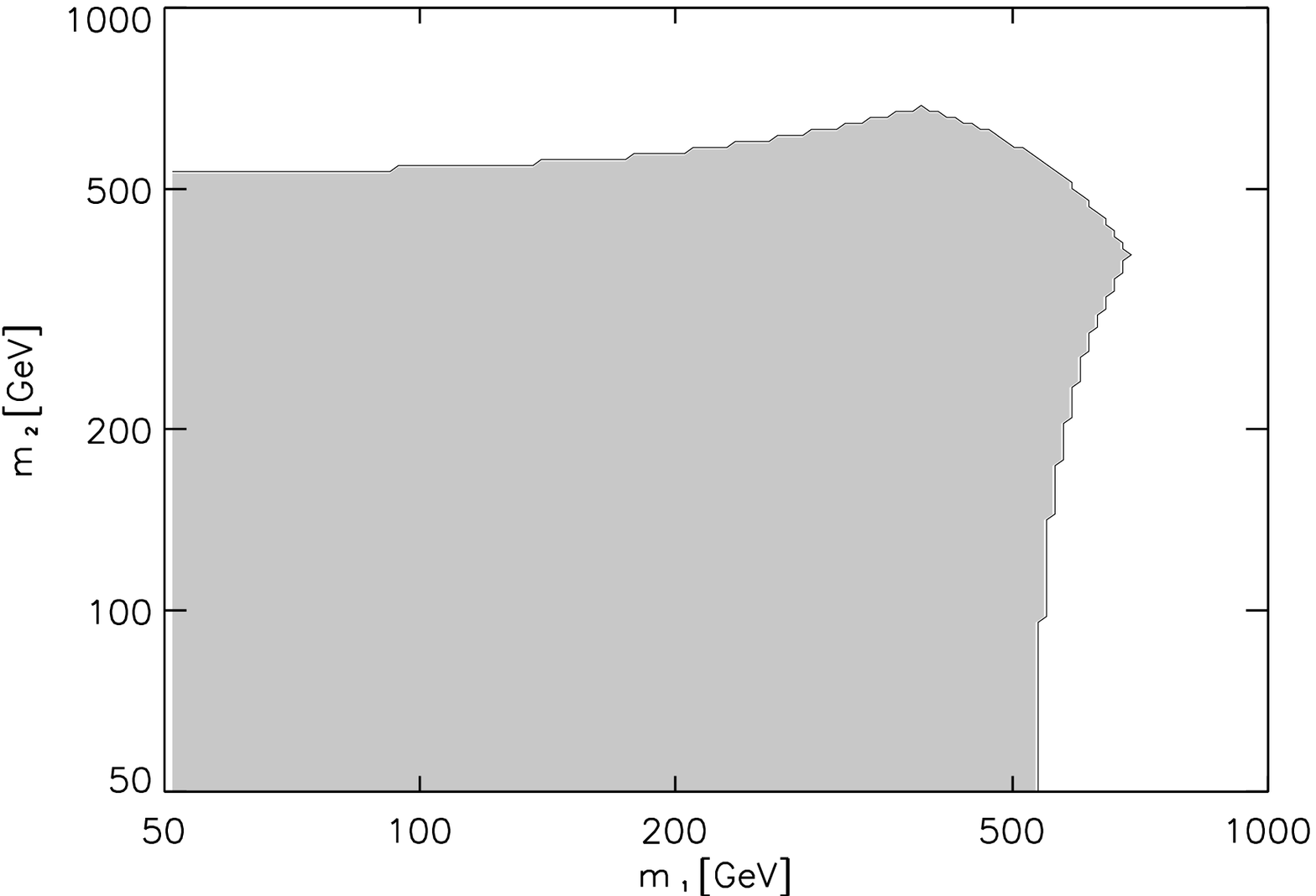,width=7.5cm,height=7.5cm}
        }
      &
      \parbox{8cm}{
        \epsfig{file=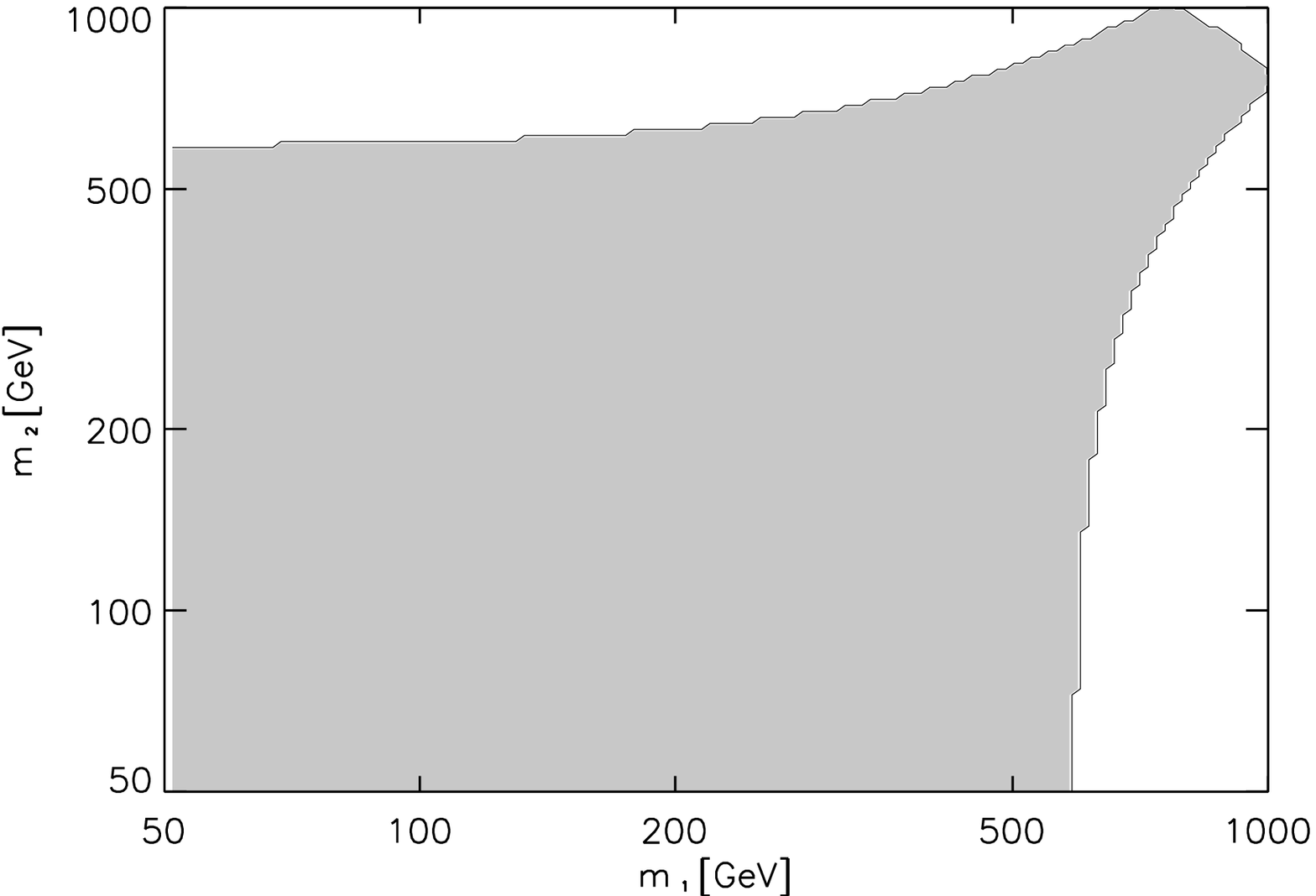,width=7.5cm,height=7.5cm}
        }
      \\
      \parbox{8cm}{
        \epsfig{file=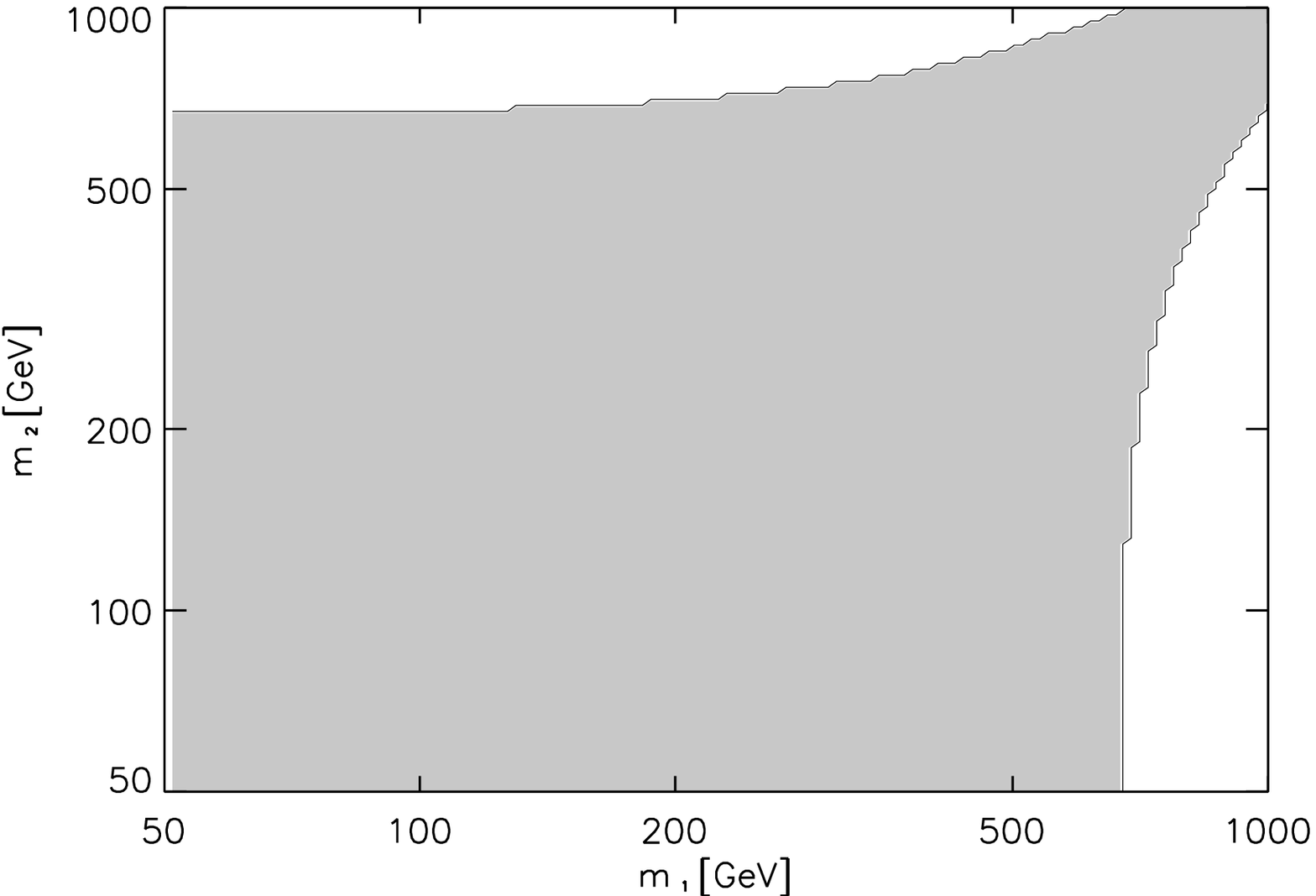,width=7.5cm,height=7.5cm}
        }
      &
      \parbox{8cm}{
        \epsfig{file=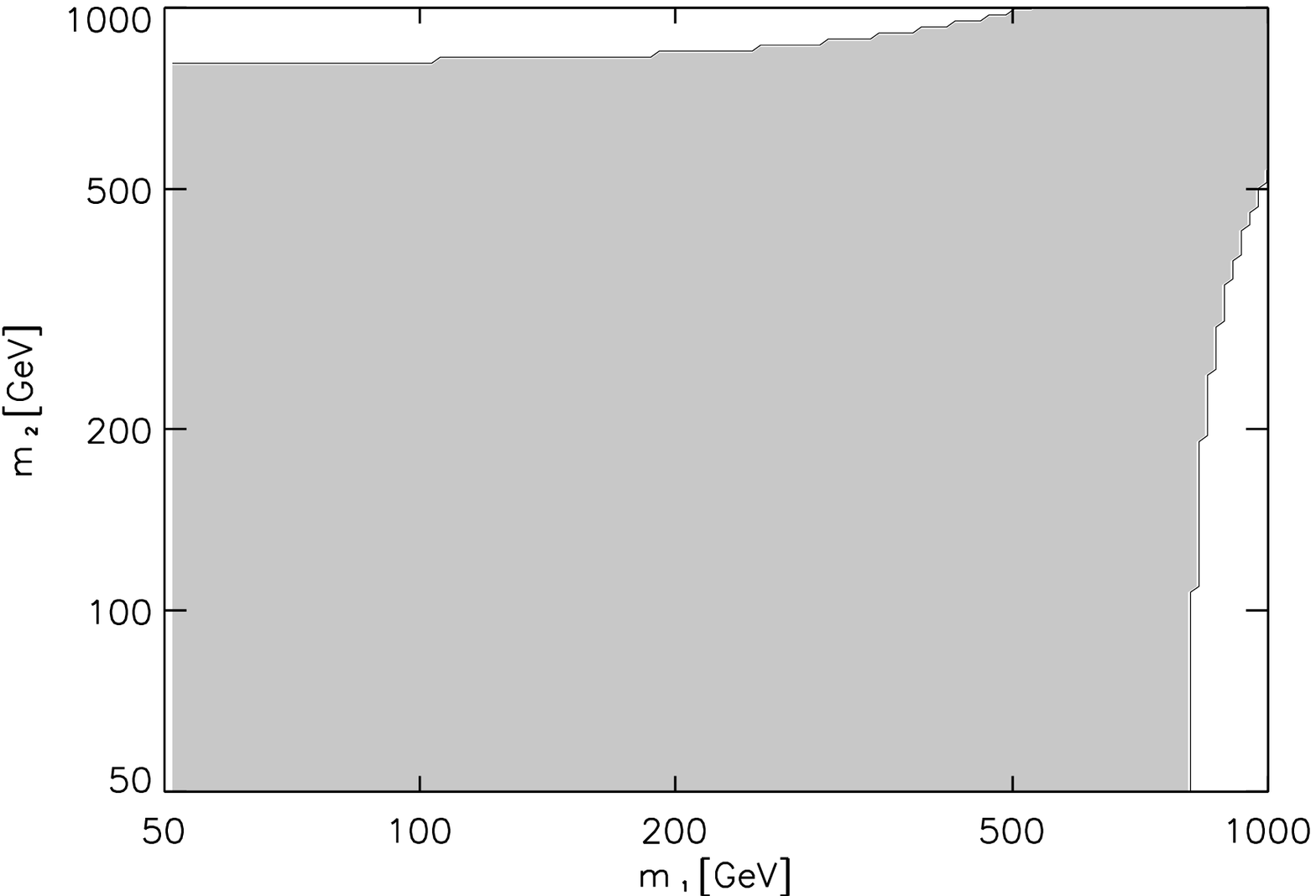,width=7.5cm,height=7.5cm}
        }
    \end{tabular}
  \end{center}
\bigskip
    \centerline{\Large\bf Fig.~3d}
\end{figure}

\begin{figure}[p]
  \begin{center}
    \leavevmode
    \epsfig{file=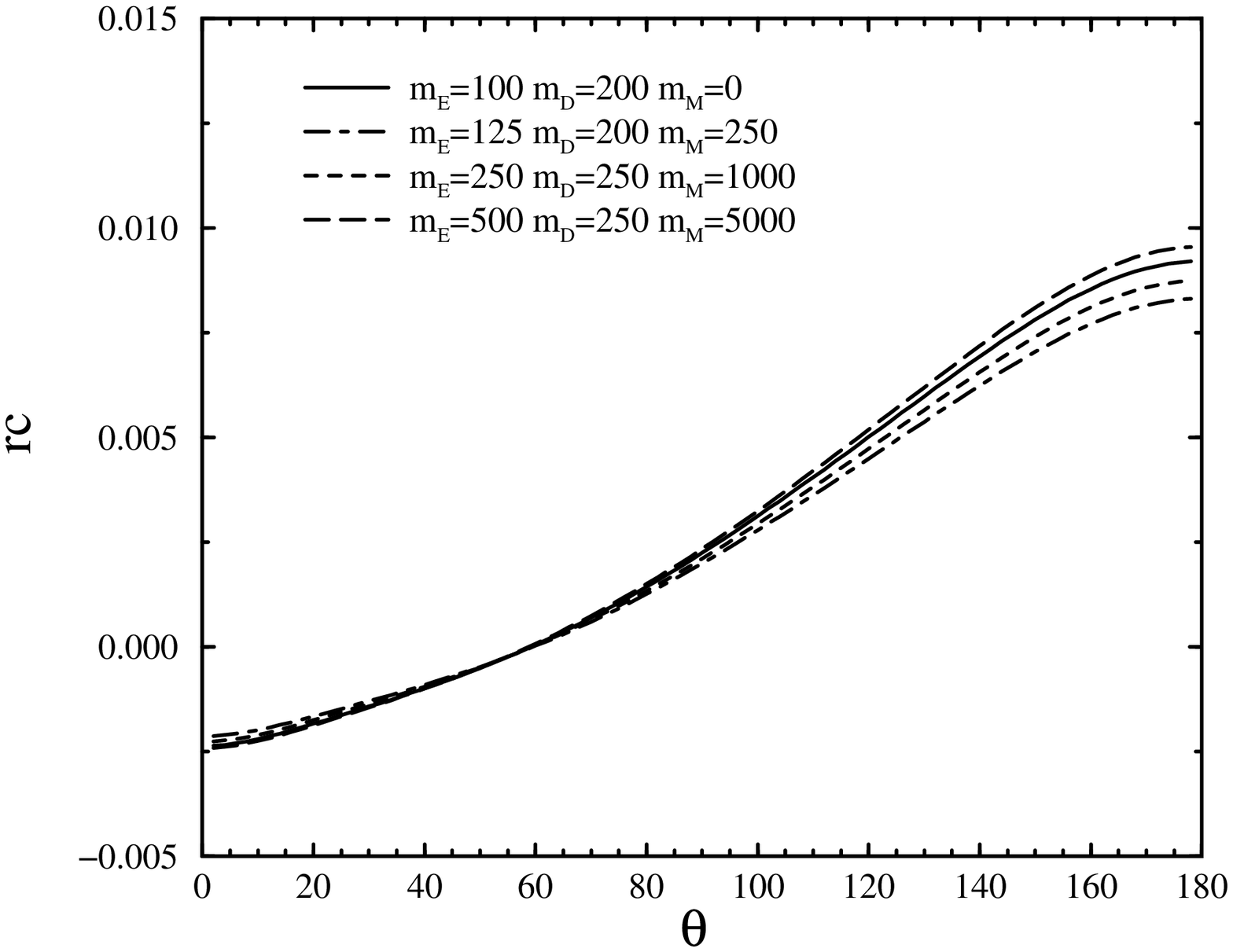,width=9.5cm,height=9.5cm}
    \centerline{\Large\bf Fig.~4}
%
    \epsfig{file=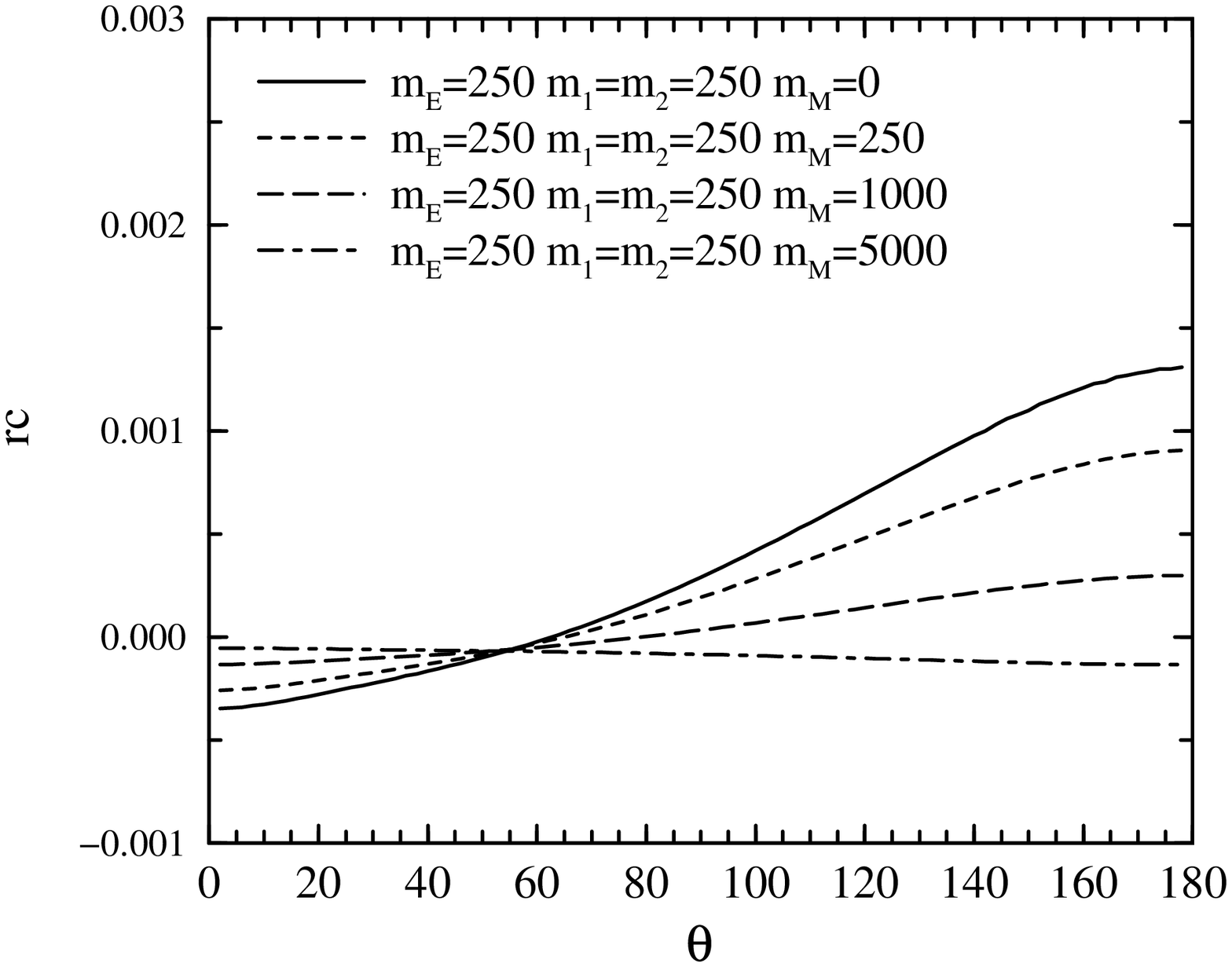,width=9.5cm,height=9.5cm}
    \centerline{\Large\bf Fig.~5}
  \end{center}
\end{figure}

\begin{figure}[p]
  \begin{center}
    \leavevmode
    \epsfig{file=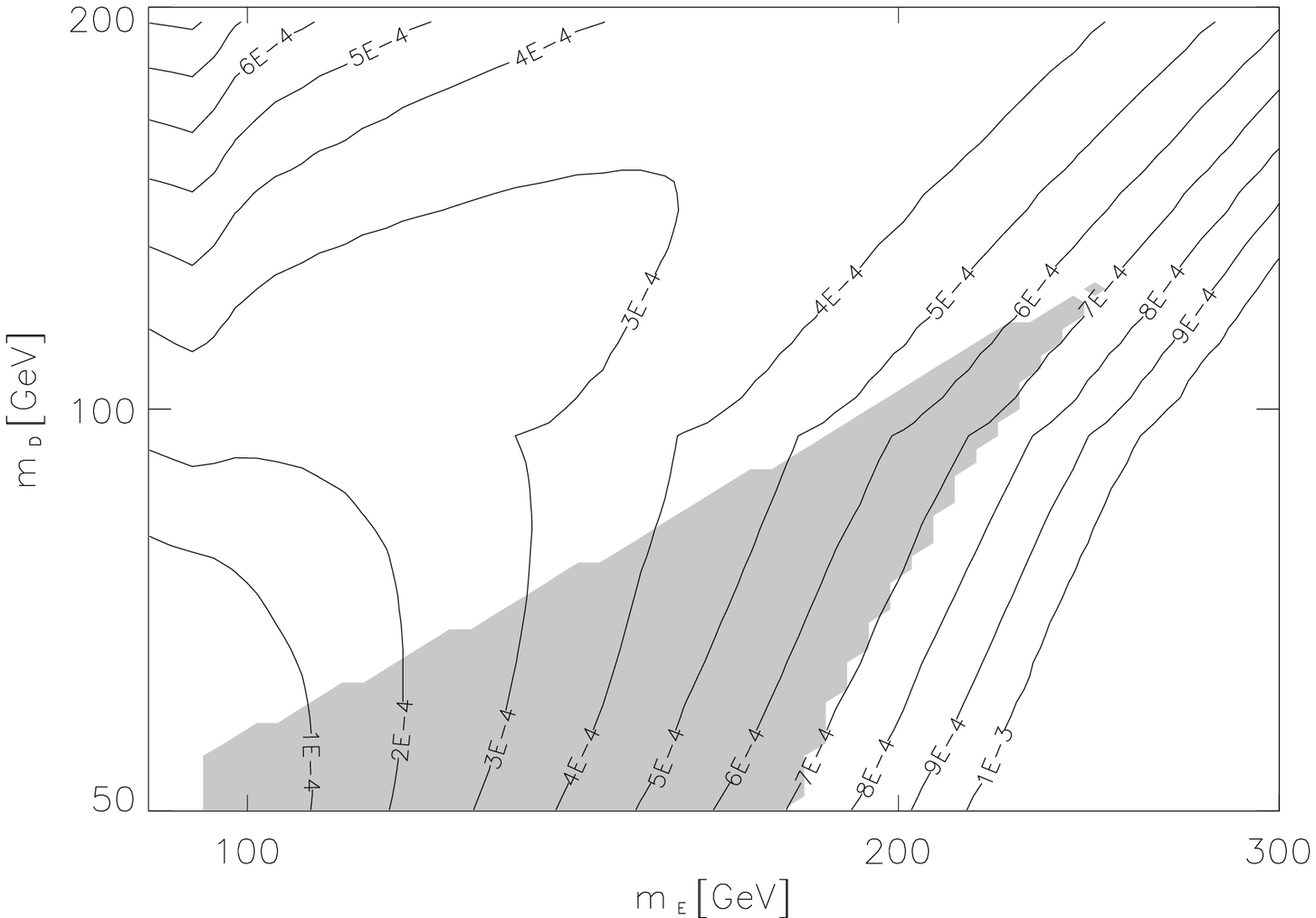,width=9.5cm,height=9.5cm}
    \centerline{\Large\bf Fig.~6a}
%
    \epsfig{file=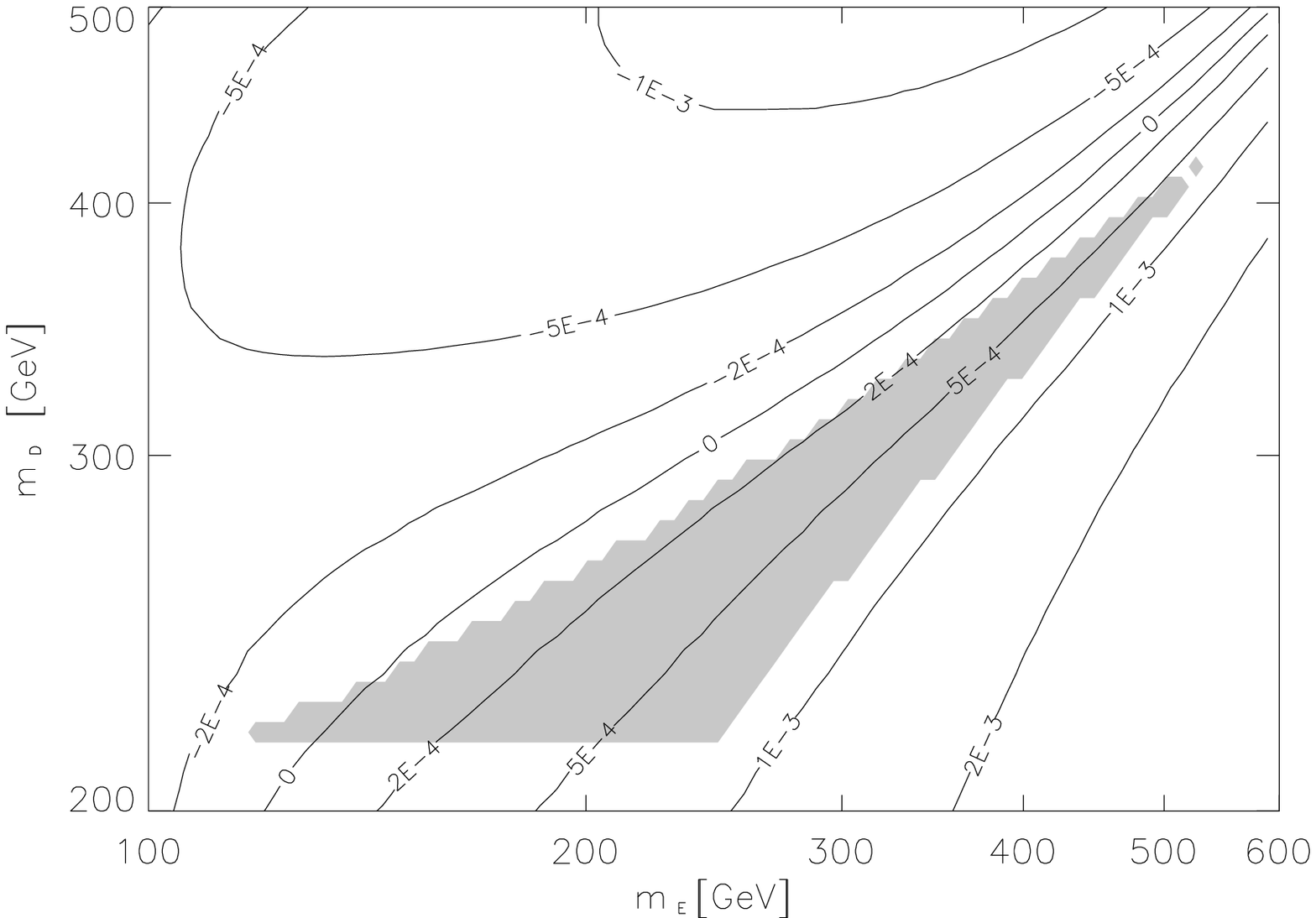,width=9.5cm,height=9.5cm}
    \centerline{\Large\bf Fig.~6b}
  \end{center}
\end{figure}

\begin{figure}[p]
  \begin{center}
    \leavevmode
    \epsfig{file=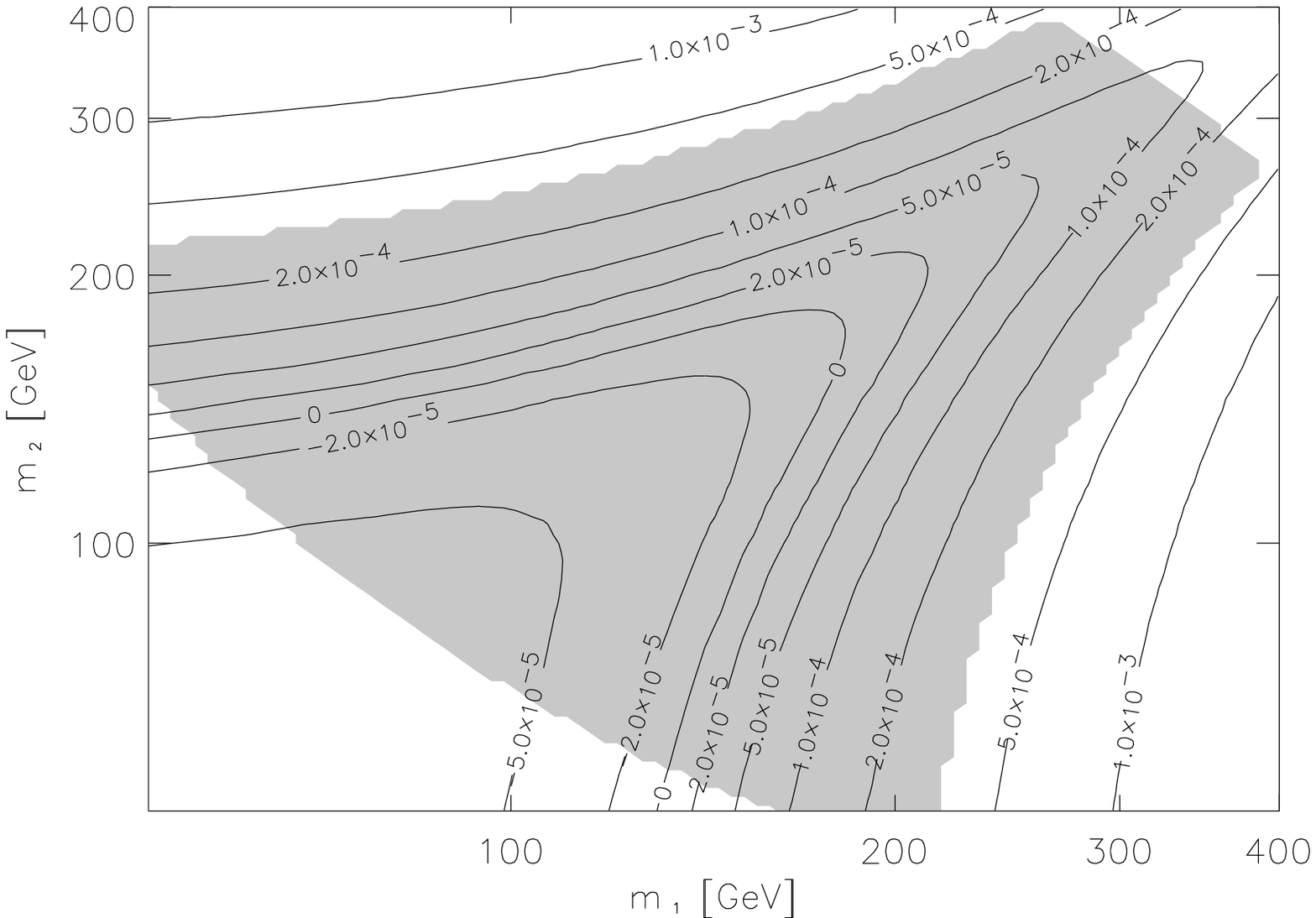,width=9.5cm,height=9.5cm}
    \centerline{\Large\bf Fig.~7a}
%
    \epsfig{file=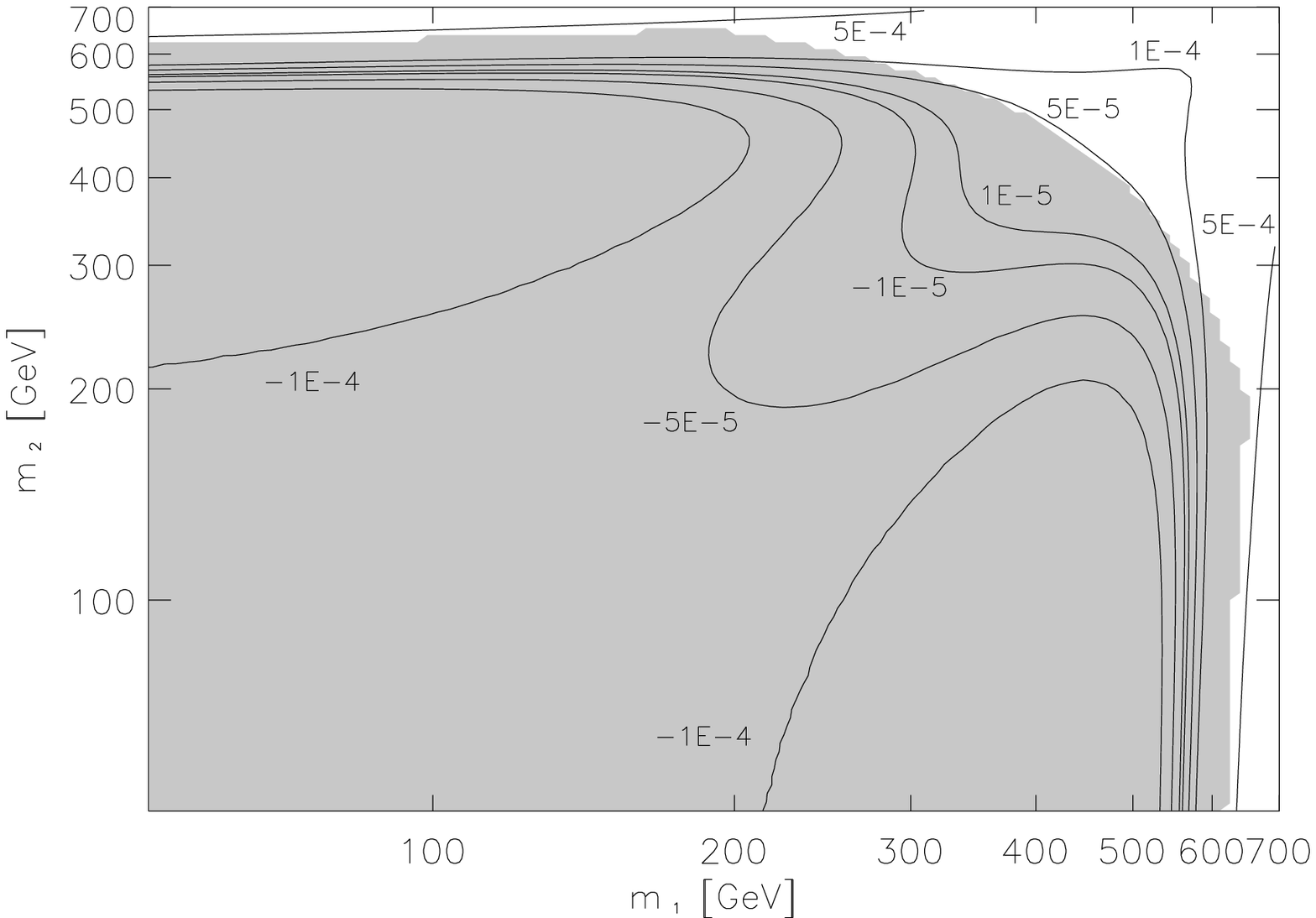,width=9.5cm,height=9.5cm}
    \centerline{\Large\bf Fig.~7b}
  \end{center}
\end{figure}

\begin{figure}[p]
  \begin{center}
    \leavevmode
    \epsfig{file=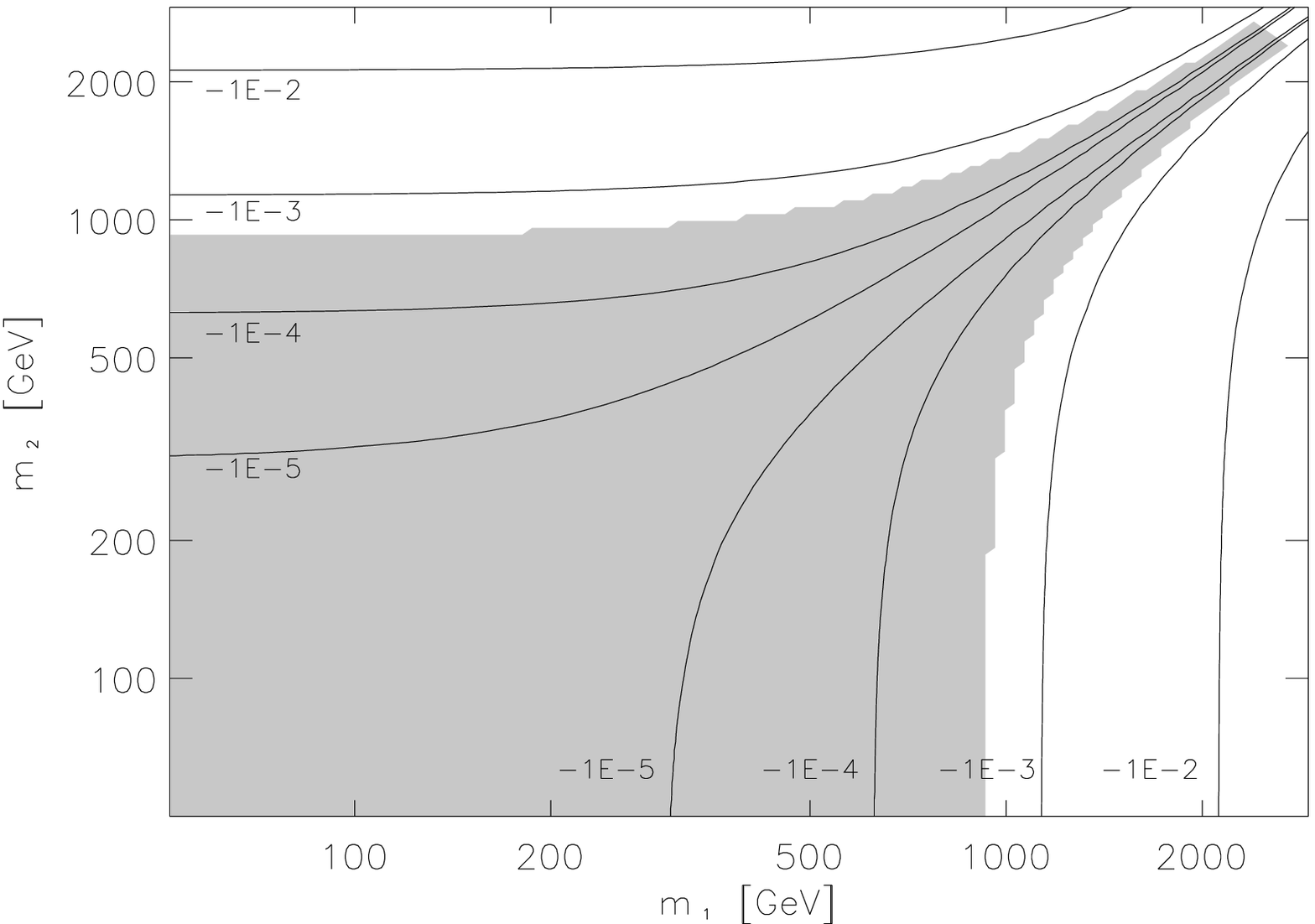,width=9.5cm,height=9.5cm}
    \centerline{\Large\bf Fig.~7c}
  \end{center}
\end{figure}
\end{document}